\documentclass[a4paper,11pt]{article}
\usepackage[T1]{fontenc}
\pdfoutput=1

\usepackage{jheppub}
\usepackage{macro}
\usepackage{appendix}
\usepackage{comment}
\usepackage{array}
\usepackage{yfonts}
\usepackage{comment}
\usepackage{amsthm}
\usepackage{tikz-cd}
\usepackage{amsmath,amssymb}
\usepackage{tikz}
\usepackage{pgfkeys}
\usepackage{tensor}
\usepackage{pgfopts}
\usepackage{xcolor}
\usepackage{young}

\usepackage{natbib}

\usepackage{youngtab}
\usepackage{ytableau}
\usepackage{titlesec}
\usetikzlibrary{decorations.pathreplacing}

\usepackage{soul}

\usetikzlibrary{matrix,calc,positioning,decorations.markings,decorations.pathmorphing,decorations.pathreplacing
}%tikzmark,
\usetikzlibrary{arrows,cd,shapes}

\def\a{\alpha} 
\def\b{\beta}

\def\d{\delta} 
\def\D{\Delta}
\def\e{\epsilon} 
\def\ve{\varepsilon}

\def\k{\kappa}

\def\m{\mu}
\def\n{\nu}

\def\S{\Sigma}

\def\f{\phi}

\def\vf{\varphi}
\def\O{\Omega}
\def\o{\omega}

\def\cB{{\cal B}}

\def\cI{{\cal I}}
\def\cJ{{\cal J}}

\def\cO{{\cal O}}

\def\hb{\hat b}
\def\he{ \hat \e}

\def\ep{\e^{\prime}}

\def\p{\prime}

\def\pr{\partial}
\title{\textbf {Higher-order $p$-form asymptotic symmetries in $D=p+2$}}

\author[a,b]{Matteo Romoli}
\author[a,b]{Federico Manzoni}

\affiliation[a]{Mathematics and Physics department, Roma Tre, Via della Vasca Navale 84, Rome, Italy}
\affiliation[b]{INFN Roma Tre Section, Physics department, Via della Vasca Navale 84, Rome, Italy}
\emailAdd{federico.manzoni@uniroma3.it, matteo.romoli@uniroma3.it}

\abstract{We investigate higher-order asymptotic symmetries for a $p$-form gauge field in  $(p+2)$-dimensional Minkowski spacetime, where Hodge duality with a scalar holds. Employing symplectic renormalization, we identify $N+1$ asymptotic charges, with each charge being parametrised by a function of the angular variables. By means of the Hodge decomposition, these charges share the same formal structure independently from $p$ and are manifestly dual to a scalar charge. We work in Lorenz gauge, therefore the gauge parameters require a radial expansion involving logarithmic (subleading) terms to ensure nontrivial angular dependence at leading order. At the same time we assume a power expansion for the field strength, allowing logarithms in the gauge field expansions within pure gauge sectors. We compute the charge algebra, which turn out to be abelian up to possible central extension due to mixed electric/magnetic charge sector and/or unexploited ambiguities during symplectic renormalization.}
\begin{document} 
	\maketitle
	
\section{Introduction}
    The field of asymptotic symmetries is an active research area born with a series of pioneering works, on asymptotically flat spacetime, of H.\ Bondi, M.\ G.\ Van der Burg, A.\ W.\ Metzner and R.\ K.\ Sachs \cite{Bondi:1962px,Sachs:1962wk,PhysRev.128.2851}. In recent decades, it has found applications within the realm, on flat spacetime, of gauge theories, such as electrodynamics  \cite{He:2014cra,Pasterski:2015zua}, Yang-Mills theory \cite{Strominger_2014,He:2015zea}, $p$-forms theories \cite{Afshar:2018apx,manz2,manz3,Francia:2018jtb,Campiglia:2018see,Henneaux_2019}, higher-spin theories \cite{Campoleoni:2017mbt,Campoleoni:2018uib} and mixed symmetry tensor theories \cite{manz4}, thanks to the established connection with soft theorems \cite{Weinberg:1964ew,Bork:2015fla,Strominger:2014pwa,He_2014,Larkoski_2014,Sen:2017xjn} and memory effects \cite{Afshar_2019,Jokela:2019apz,Ball_2019,PhysRevLett.117.061102}. Asymptotic symmetries are also studied in arbitrary spacetime dimension \cite{Campoleoni:2019ptc,Campoleoni:2020ejn,He:2019jjk,He:2019pll,Henneaux:2019yqq}, different backgrounds \cite{Henneaux:1985tv, Brown:1986nw, Esmaeili:2019mbw, Esmaeili:2021szb, Campoleoni:2023eqp} and in more general contexts such as holographic correspondence \cite{Barnich:2010eb}, duality in gauge theories \cite{manz2,manz3,manz4}, convolutional double copy \cite{Ferrero:2024eva} and gravisolitons \cite{Manzoni:2021dij,Manzoni:2024agc}. For a beautiful self-contained review see \cite{strominger2018lectures}. 

    Regardless of the specific theory, the program of asymptotic symmetries consists of finding the residual gauge symmetries that preserve the boundary conditions (the chosen falloffs) and act non-trivially on asymptotic field configurations\footnote{The word ``asymptotic'' should not necessarily be taken in the sense of infinite distance but rather of behaviour very close to a codimension two spacelike manifold embedded in spacetime.}. In this paper, we are interested in discussing boundary conditions at null infinity and, in	particular, we focus our analysis on $\cI^+$, which we reach in the limit $r \rightarrow \infty$ at fixed retarded time $u = t-r$. The falloffs are determined by asking finiteness of all relevant physical quantities, such as the energy flux, as well as the presence of all physically reasonable solutions of the theory.

    Our focus is on $p$-forms gauge fields that emerge quite naturally in many physical contexts: in string theory, $Dp$-branes can be introduced as sources carrying the charge of $(p+1)$-form and $p$-form field are part of the string spectrum \cite{green_schwarz_witten_2012}; in supersymmetric theories, multiplets of supersymmetry algebra generically contain $p$-forms \cite{cecotti_2015}. $p$-forms are also introduced after Kaluza-Klain reduction in supergravity and in the context of AdS/CFT correspondence \cite{lYi:1998icc} and they have recently received interest in the context of celestial holography \cite{Donnay:2022ijr}. We fix the dimension to be $D=p+2$, hence representation duality with a scalar holds. On the one hand, we motivate this choice observing that scalar theories are the simplest models in physics and therefore they might give insights on the physical interpretation of the asymptotic charges. Furthermore, scalar falloffs are generally  accepted to be the standard radiation falloffs, while for other theories several generalisations were proposed. In fact, from the duality we derive the field strength falloffs and, by studying the equation of motion and the gauge fixing condition, we determine the gauge field falloffs in \autoref{sec3}. 
	
    Working in Lorenz gauge, parameters must admit logarithmic (subleading) terms in their expansion  \cite{He:2019jjk,He:2019pll,Campoleoni:2019ptc,manz2,Hirai:2018ijc,Romoli:2024hlc,manz5}. This is required to obtain terms with arbitrary and nontrivial angular dependence and therefore a nontrivial asymptotic charge associated to residual gauge transformations. For what concerns the gauge-field components, we consider logarithms only in pure gauge sectors, so that the field strength admits a simple power-law expansion. 
	
	Moving around the ideas that led to the formulation of superrotations \cite{Barnich:2009se,Barnich:2010ojg,Campiglia:2014yka}, asymptotic symmetries have been generalised to higher orders, where we define the order of an asymptotic symmetry in relation to the radial power of the corresponding asymptotic charge (before renormalization). For instance, supertranslations in $D=4$ are of order $\cO(1)$, while superrotations are of order $\cO(r)$. In this respect, the idea at the basis of the infrared triangle is that the semiclassical Ward Identity stemming from an higher-order asymptotic symmetry is equivalent to a certain subleading soft theorem. We can find analyses of $\cO(r)$ asymptotic symmetries in electromagnetism \cite{Campiglia:2016hvg}, recently generalised to arbitrary $\cO(r^N)$ in \cite{Peraza:2023ivy} and deeply studied in two-form gauge theories \cite{Romoli:2024hlc} by one of the authors of the present work. Moreover, we stress that $\cO(r^N)$ asymptotic symmetry was found also in \cite{manz2} as magnetic-like counterpart of a $\cO(r^{-M})$ electric-like asymptotic charge with a relation between $N$ and $M$ directly inherited by the representation (Young diagrams) duality of $p$-form gauge theories. The $p=1$ non-abelian case was studied at $\cO(r)$ in \cite{Campiglia:2021oqz}, and recently extended to arbitrary
	higher order in \cite{Nagy:2024dme,Nagy:2024jua}. For gravity, $\cO(r^2)$ asymptotic symmetries were studied in \cite{Campiglia:2016efb}, exploring the connection with sub-subleading soft graviton theorem, and in the self-dual sector in \cite{Nagy:2022xxs}. Our analysis fits into the broader recent perspective in which symmetries are studied: asymptotic symmetries, generalized symmetries, non-invertible symmetries and the possible interplay between them in gravity and gauge theory \cite{Gaiotto:2014kfa,Hinterbichler:2022agn,Benedetti:2021lxj, Cheung:2024ypq,Benedetti:2023ipt,manz3,Afshar:2018apx,Lake:2018dqm,Hofman_2018}. 
    
    To compute the corresponding $\cO(r^N)$ asymptotic charge, we employ the Hodge decomposition on a $p$-dimensional sphere to decompose the gauge parameter and the relevant field components. The asymptotic charge, that has the same formal expression regardless the form degree, is then expressed in term of the free Cauchy data that must be distinguished for even and odd degree forms. In view of the scalar-form duality in $D=p+2$, it is interesting to note that we compute and renormalize charges also in odd dimension. This can be useful to better understand the asymptotic charges of scalars that in literature have been calculated mainly in even dimension \cite{Campiglia_20182,Campiglia_2018,Francia:2018jtb}.

	In the case of higher-order asymptotic symmetries the relevant issue of the divergences of the charges has to be taken into account. These divergences can be understood as ambiguities of the presymplectic potential and therefore renormalized following \cite{Freidel:2019ohg}. We apply this procedure to our asymptotic charges, aiming to cancel all the divergences while leaving unaltered the finite parts.  Using the ambiguities in the definition of the presymplectic potential we can introduce appropriate counterterms that cancel out the divergent parts of the charges. Assuming our set up, the procedure is completely general and lies in the possibility of a general expansion of the presymplectic potential $\theta^t$ (expansion \eqref{exp1} in terms of $t$) and of its finite part $\theta^t_{(0)}$ (expansion \eqref{exp2} in terms of $u$).

    In \autoref{sec2}, we review the Hodge theory, the covariant phase space formalism and the symplectic renormalization procedure. The first is instrumental to reduce the $p$-form case to a scalar one, obtaining a general $p$-independent expression for the asymptotic charges (up to some overall factors), derived in \autoref{sec3}. The second and third ones are used to introduce the ambiguities by means of which, in \autoref{sec4}, we renormalize the divergent terms in the asymptotic charge. Finally in \autoref{sec5} we compute the charge algebra and discuss possible central extensions. 

%%%%%%%%%%%%%%%%%%%%%%%%%%%%%%%%%%%%%%%%%%%%%%%%%%%%%%%%%%%%%%%%%%%%%%%%%%%%%%%%%%%%%%%%%%%%%%%%%%%%%%%%%%%%%%%%%%%%%%%%%%%%%%%%%%%%%%%%%%%%%%%%%%%%%%%%%%%%%%%%%%%%%%%%%%%%%%%%%%%%%%%%%%%%%%%%%%%%%%%%%%%%%%%%%%%%%%%%%%%%%%%%%%%%%%%%%%%%%%%%%%%%%%%%%%%%%%%%%%%%%%%%%%%%%%%%%%%%%%%%%%%%%%%%%%%%%%%%%%%%%%%%%%%%%%%%%%%%%%%%%%%%%%%%   
    
\section{Preliminaries}\label{sec2}
	
	\subsection{Notation and conventions}
	We consider $p$-forms in dimension $D = p+2$ and adopt the mostly plus convention for the metric. For the discussion of asymptotic parameter, we are going to use retarded Bondi coordinates $(u, r, \{x^i\})$, with $r$ being the
	radial coordinate, $u := t - r$ the retarded time and $\{x^i\}$, with $i \in [1,D-2]$, the angular coordinates. In these coordinates, the Minkowski spacetime line element is given by 
	\begin{equation}
		ds^2=-du^2-2dudr+r^2\gamma_{ij}dx^idx^j\, ,
	\end{equation}
	where $\gamma_{ij}$ is the unit $(D-2)$-sphere metric.
	Metric tensor components are
	\begin{equation}
		g_{\m \n} =\begin{bmatrix}
			-1 &-1 & 0\\
			-1 & 0 & 0\\
			0 & 0 & r^2\gamma_{ij}
		\end{bmatrix}, \ \ \ 
		g^{\m \n} =\begin{bmatrix}
			0 & -1 & 0\\
			-1 & 1 & 0\\
			0 & 0 & \frac{1}{r^2}\gamma_{ij}^{-1}
		\end{bmatrix},
	\end{equation}
	and the non-vanishing Christoffel symbols are
	\begin{equation}
		\Gamma^i_{jr}=\Gamma^i_{rj}=\frac{1}{r}\delta^i_j\,, \ \ \ \Gamma^u_{ij}=-\Gamma^r_{ij}=r\gamma_{ij}\,, \ \ \ \Gamma^k_{ij}=\frac{1}{2}\gamma^{kl}[-\pr_l\gamma_{ij}+\pr_j\gamma_{li}+\pr_i\gamma_{jl}] \, .
		\label{cribon}
	\end{equation}
	
	The covariant derivative w.r.t.\ $\gamma_{ij}$ is denoted with $D_i$ and the $(D-2)$-sphere Laplace-Beltrami operator is $\Delta:=D_iD^i$. Similarly, we use $\nabla_{\m}$ for the covariant derivative associates to the Riemannian structure given by the metric tensor $g_{\m \n}$ and $\Box:=\nabla_{\m}\nabla^{\m}$ for the d’Alembert operator. 
	
	Future null infinity $\cI^+$ is reached in the limit $r \rightarrow \infty $ (or, equivalently, $t \rightarrow \infty $) at fixed $u$: it is a null hypersurface homeomorphic to $\mathbb{R} \times S_u^{D-2}$. In the limit $u \rightarrow \pm \infty$ we move towards the ``endpoints'' of null infinity, dubbed $\cI^+_{\pm}$. 
    
    We are interested in expanding fields and parameters near $\cI^+$. The most general radial expansion we consider is
	\begin{equation}
		\varphi=\sum_{n_1}\frac{\varphi^{(n_1)}}{r^{n_1}}+\sum_{n_2}\hat{\varphi}^{(n_2)}\frac{\ln {(r)} }{r^{n_2}} \, ,
	\end{equation}
	where the coefficients of the radial expansion are always denoted with a superscript “$(n)$” and, in particular, we use the hat for the logarithmic coefficients. When discussing charges and renormalization we employ $(t, u, \{x^i\})$ coordinates, since we are interested in the large $t$-behaviour around null infinity. The expansion coefficients are
	denoted with a subscript “$(n)$” and 
	\begin{equation}
		\varphi=\sum_{n_1}{t^{n_1}}{\varphi_{(n_1)}}+\ln{(t)} \sum_{n_2}{t^{n_2}}{\hat{\varphi}_{(n_2)}} \, .
	\end{equation}
    Let us remark that $n_1,n_2$ are integers for even values of $p$, while are semi-integers for odd $p$. 
    
	To describe the behaviour near $\cI^+_-$ we consider only power expansions that we denote as
	\begin{equation}
		\varphi^{(n)}=\sum_mu^m\varphi^{(n,m)} \,,\qquad \varphi_{(n)}=\sum_mu^m\varphi_{(n,m)} \, .
	\end{equation}
	For the gauge parameter, this expansion is valid for all values of $u$ and not only for $u \rightarrow -\infty$, but we use the same notation, always specifying its meaning. In the $u$-expansions we consider, $m$ takes always integer values. 

%%%%%%%%%%%%%%%%%%%%%%%%%%%%%%%%%%%%%%%%%%%%%%%%%%%%%%%%%%%%%%%%%%%%%%%%%%%%%%%%%%%%%%%%%%%%%%%%%%%%%%%%%%%%%%%%%%%%%%%%%%%%%%%%%%%%%%%%%%%%%%%%%%%%%%%%%%%%%%%%%%%%%%%%%%%%%%%%%%%%%%%%%%%%%%%%%%%%%%%%%%%%%%%%%%%%%%%%%%%%%%%%%%%%%%%%%%%%%%%%%%%%%%

\subsection{The Hodge decomposition and duality}
	Given a differential manifold $(M, \mathcal{A})$ with maximal atlas $\mathcal{A}$, we can consider a cochain complex given by differential forms and with differential given by the de Rham differential $d$ which is a map from the module of $k$-forms to the module of $(k+1)$-forms
	%\begin{equation}
	%	d : \Omega^{k}(M) \rightarrow \Omega^{k+1}(M) \, .
	%\end{equation} 
    which is the unique family of maps that generalise the standard differential of a $C^{\infty}(M)$ function thanks to the Cartan theorem.
	This complex is
	the de Rham complex $C_{\text{dR}}^{*}$
	\begin{equation}
		{\displaystyle 0 {\stackrel {}{\to }}\ \Omega ^{0}(M)\ {\stackrel {d}{\to }}\ \Omega ^{1}(M)\ {\stackrel {d}{\to }}\ \Omega ^{2}(M) \ {\stackrel {d}{\to }}\ \cdots {\stackrel {d}{\to }}\ \Omega ^{D-1}(M)\ {\stackrel {d}{\to }}\ \Omega ^{D}(M)\ {\stackrel {}{\to }}\ 0} \, .
	\end{equation}
	Let us assume that our differential manifold $M$ is compact, orientable and endowed with a pseudo-riemannian structure we can define the Hodge star linear operator, which maps $k$-forms to $(D - k)$-forms.
    It has the following property, which defines it completely 
	\begin{equation}
		{\displaystyle \a \wedge ({\star }\b )=\langle \a ,\b \rangle \,\omega_{\mathrm{vol}} } \qquad \a ,\b \in \Omega^k(M) \, .
	\end{equation}
	where $\wedge$ is the wedge product of forms, $\langle .,.\rangle$ is the inner product induced by the metric tensor and $\omega_{\mathrm{vol}}\equiv\star 1$ is the volume form. Modules of form are vector spaces, hence the Hodge star is a one-to-one mapping since the dimensions of these spaces are the binomial coefficients ${\displaystyle {\tbinom {D}{k}}={\tbinom {D}{D-k}}}$. For us, the relevant case is that of $k=1$ and $D=p+2$. Choosing the 1-form as the field strength of a scalar $\phi$ and the $(p+1)$-form as the field strength of a $p$-form $\mathcal{B}$ we obtain the relation
    \begin{equation}
        d\mathcal{B}=\star d \phi \, .
    \end{equation}
    The above relation is the covariant version at the field strengths level of a duality that emerges, through the use of Young diagrams, from the theory of representations of the little group $\mathrm{SO}(D-2)$ and which therefore leads to an on-shell physical duality \cite{Hull:2001iu,Medeiros_2003,hamermesh1989group}. In our case, where $D=p+2$, this duality leads to the possibility of equivalently describing a $p$-form theory via a scalar theory.
    
  Through the Hodge operator, we can define the codifferential 
  \begin{equation}
      d^{\dagger}:=(-1)^{D(k+1)+1}s\ {\star }d{\star }=(-1)^{k}\,{\star }^{-1}d{\star }
  \end{equation}
  where $s$ is the sign of the determinant of the metric tensor. 
  Then
	\begin{equation}
		d^{\dagger} :\Omega ^{k}(M)\to \Omega ^{k-1}(M) \, ,
	\end{equation}
	and satisfy $d^{\dagger} \circ d^{\dagger}=0$.
	At this point we can define the Laplace-de Rham operator
	\begin{equation}
		\Delta_{\text{dR}}:=(d^{\dagger}+d)^2=d^{\dagger}d+dd^{\dagger} \, ;
	\end{equation}
	by definition, a form on $M$ is harmonic if its Laplacian is zero:
	\begin{equation}
		{\mathcal {H}}_{\Delta_{\text{dR}} }^{k}(M):=\{\gamma \in \Omega ^{k}(M)\mid \Delta_{\text{dR}} \gamma =0\} \, .
	\end{equation}
	The Hodge operator maps harmonic forms to harmonic forms. As a consequence of the Hodge theory, the de Rham cohomology is isomorphic to the space of harmonic $k$-forms. The Hodge decomposition theorem states that any differential form $\omega$ on a closed Riemannian manifold can be written as a sum of three parts
	\begin{equation} \omega =d\a +d^{\dagger} \b +\gamma \, ,  
	\end{equation}
	in which $\gamma$ is harmonic. 
    
	In what follows, we consider the $p$-dimensional sphere (which is a compact, orientable riemannian manifold) and the Hodge decomposition is employed for the gauge parameter with only angular indexes $\e_{i_1 \dots i_{p-1}}$ and for the field components with only $p-1$ angular indexes. In both cases we have to deal with a $(p-1)$-form, hence the Hodge decomposition reduces to 
	\begin{equation}\label{hodge}
        \omega=d\a +d^{\dagger} \b \quad \o \in \O^{p-1}(S^p) \, ,  
	\end{equation}
	since 
	\begin{equation}
		\mathcal {H}_{\Delta_{\text{dR}} }^{p-1}(S^p) \cong H^{p-1}_{\text{dR}}(S^p; \mathbb{R})=0 \, ,
	\end{equation}
	since the $p$-sphere has non-trivial cohomology only in degrees 0 and $p$. In the specific case of a 1-form $V$ on the 2-sphere the Hodge decomposition reads 
	\begin{equation}
		V=d\a+d^{\dagger}\b \qquad \a \in \Omega^0(S^2), \ \b \in \Omega^2(S^2) \, ;
		\label{Hodge1}
	\end{equation}
	which, in coordinates, is given by 
	\begin{equation}
		V_i=D_i\a+D^j\b_{ij}=D_i\a+\ve_{ij}D^j\b \, ,
		\label{Helmoltz}
	\end{equation}
	where $\ve_{ij}$ is the two-sphere Levi-Civita tensor, and $\b_{ij}=\ve_{ij}\b$ since it is a top form. The writing \eqref{Helmoltz} is also known as the Helmholtz decomposition. For a $(p-1)$-form $V$ on the $p$-sphere, equation \eqref{Helmoltz} is generalised to
	\begin{equation}
		V_{i_1\dots i_{p-1}} = D_{[i_1} X_{i_2\dots i_{p-1}]} +  \ve_{i_1\dots i_{p}}D^{i_p}Y \, .
	\end{equation}
	For a $p$-form, since it is a top form, we can write instead
	\begin{equation}
		V_{i_1\dots i_{p}} = \ve_{i_1 \dots i_p}V \, .
	\end{equation}
    
%%%%%%%%%%%%%%%%%%%%%%%%%%%%%%%%%%%%%%%%%%%%%%%%%%%%%%%%%%%%%%%%%%%%%%%%%%%%%%%%%%%%%%%%%%%%%%%%%%%%%%%%%%%%%%%%%%%%%%%%%%%%%%%%%%%%%%%%%%%%%%%%%%%%%%%%%%%%%%%%%%%%%%%%%%%%%%%%%%%%%%%%%%%%%%%%%%%%%%%%%%%%%%%%%%%%%%%%%%%%%%%%%%%%%%%%%%%%%%%%%%%%%%

\subsection{Covariant phase space formalism}
	The most important tool in order to understand whether an asymptotic symmetry has a physical interpretation is the asymptotic charge. We introduce this notion employing the covariant phase space formalism, whose main idea is to combine the differential calculus in both spacetime and field space to give rise to the variational bicomplex and jet bundle. 
	
	Let $(M,\boldsymbol{g})$ be a Lorentzian manifold of dimension $D$ with maximal atlas $\mathcal{A}$ and consider an arbitrary field-theory with fields denoted by ${\Phi}^I$, with $I$ denoting a totally general index structure, and Lagrangian density $\mathcal{L}$. The variation of the Lagrangian density can be expressed as 
	\begin{equation}\label{varlagrgen}
		\delta \mathcal{L}=\mathrm{EOM}_I\delta \Phi^I+d{\theta}
	\end{equation}
	where the first term contains the Euler-Lagrange equations of motion ($I$ indicates an arbitrary indices structure) while the second one is a boundary term called the bare presymplectic potential. 

    Explicitly, the variational operator $\delta$ is defined as
	\begin{equation}
		\delta:=\delta \Phi^I\frac{\pr}{\pr \Phi^I}+\delta (\pr_{\m}\Phi^I)\frac{\pr}{\pr (\pr_{   \m}\Phi^I)}+...+\delta (\pr_{(\m}...\pr_{\n)}\Phi^I)\frac{\pr}{\pr(\pr_{(\m}...\pr_{\n)}\Phi^I)} \, .
	\end{equation}
    Using the convention according to which each field or derivative field variation is Grassmann odd, the variation operator satisfies $\delta \circ \delta=0$; this means we can construct a cohomology with $\delta$ and every field or derivative field variation is a 1-form in the jet space $J$. Merging the spacetime $M$ and the jet space $J$ we obtain the jet bundle and variational bicomplex \cite{VINOGRADOV19841, VINOGRADOV198441, Compere:2018aar}. Essentially, the variational bicomplex is a cochain complex of the differential graded algebra of exterior forms on jet manifolds of sections of a fiber bundle: lagrangians and Euler–Lagrange operators on a fiber bundle are defined as elements of this variational bicomplex while its cohomology leads to the Noether theorems. The jet bundle is locally a manifold with local coordinates $(x^{\m}, \Phi^I,g_{\m \n}, \pr_{\a}\Phi^I, \pr_{\a}g_{\m \n}, \pr_{\a}\pr_{\b}\Phi^I, \pr_{\a}\pr_{\b}g_{\m \n},...)$; we refer to a $(p,q)$-form on this variational bicomplex as a spacetime $p$-form and a field-space $q$-form.
    
	In \eqref{varlagrgen}, $\mathcal{L}$ is a $(D,0)$-form, while ${\theta}$ is a $(D-1,1)$-form. The local presymplectic form is defined as 
	\begin{equation}
		{\omega}:=\delta {\theta},
	\end{equation}
    which is a $(D - 1, 2)$-form; this local expression can be integrated on a Cauchy slice $\Sigma$ to give the presymplectic two-form
	\begin{equation}\label{formgu}
		{\Omega}=\int_{\Sigma}{\omega} \, ,
	\end{equation}
	and from its definition, under sufficiently smooth hypothesis on $\Sigma$, we have $\delta {\Omega}=0$. Let us consider a vector field $V \in \chi(J)$ such that the field space Lie derivative satisfies 
	\begin{equation}
		L_V{\omega}=0 \, .
	\end{equation}
    Given a gauge transformation
    \begin{equation}
         B \;\mapsto\; B + d\e \, ,
    \end{equation}
    this induces, in the covariant phase space formalism, a vertical vector field
     \begin{equation}
         V_{\e} \in \chi(J)
     \end{equation}
    defined by its action on the fields:
     \begin{equation}
        \delta_{V_\e} A = d\e \, .
        \end{equation}
    Using Cartan magic formula and the nilpotency of the differential we get
	\begin{equation}
		L_V{\omega}=(i_V\delta+\delta i_V){\omega}=\delta (i_V{\omega})=0 \, ,
		\label{liesimp}
	\end{equation}
	where $i_V$ is the interior product for the field space $J$. Equation \eqref{liesimp} teaches us that $i_V{\omega}$ is a cocycle. Therefore, assuming trivial cohomology\footnote{In other words, we require exactness on the space of 1-forms on $J$.} in the space of 1-forms on $J$
	\begin{equation}
		i_V{\omega}=\delta {\cJ}_V \, ,
	\end{equation}
	where the $(D-1,0)$-form $
	{\cJ}_V$ is the Noether current. %whose components are the Hodge dual of those defined in the standard way. 
	We can define a global functional as
	\begin{equation}
		Q_V:=\int_{\Sigma} {\cJ}_V \, ,
		\label{Hv}
	\end{equation}
	such that
	\begin{equation}
		i_V{\Omega}=\delta Q_V \, .
		\label{int}
	\end{equation}
    In general, the variation $\delta Q_V$ does not define an integrable charge on phase space. However, in the linearised theories considered in this work, this obstruction is absent and the charges are integrable.
	Furthermore, for a gauge symmetry the current is in general an on-shell total derivative
    \begin{equation}\label{Noether-two-form}
        \cJ_V^\m = \pr_\n \k^{[\m\n]}\,,
    \end{equation}
	where $\k$ is usually referred as Noether two-form, so that the charge can be written as a integral over $\pr \Sigma$ by means of the Stokes theorem. 

\subsection{Symplectic renormalization}\label{parsr}
The roots of symplectic renormalization can be traced back to the introduction of the
covariant symplectic structure on the space of classical solutions \cite{Zuckerman:1986vzu}. 
Soon after, the role of boundary terms was clarified and was showed that the presymplectic potential is defined only up some ambiguities \cite{Lee:1990nz,Iyer_1994}. 
The presymplectic potential admits two types of ambiguities
	\begin{equation}\label{presmb}
		\theta \sim_{\text{eq}} \theta + \delta \Xi +d\Upsilon.
	\end{equation}
	The first does not change $\omega$, since $\delta \circ \delta =0$, and corresponds to the addition of a
	boundary term to the Lagrangian $\mathcal{L} \sim_{\text{eq}} \mathcal{L} + d\Xi$. The second modifies $\omega \sim_{\text{eq}} \omega + d \delta \Upsilon$  but
	does not affect $d\theta$, since $d \circ d=0$.
Symplectic renormalization refers to the procedure of modifying the
presymplectic potential by exploiting the ambiguities in its definition so
that the resulting symplectic form is well defined. Concretely, one starts from a local
presymplectic current $\omega$ derived from a Lagrangian $\mathcal{L}$,
and observes that the integral \eqref{formgu}
may diverge or fail to be independent of the choice of Cauchy surface.
The key insight is that the ambiguities \eqref{presmb} can be exploited to cure these problems.
By choosing ambiguities appropriately, often guided by boundary conditions or
asymptotic symmetries, one obtains a renormalized symplectic form which is finite and well-defined on the physical phase space.  

Methodologically, symplectic renormalization closely parallels
holographic renormalization \cite{Henningson_1998,de_Haro_2001,Skenderis_2002}. One introduces a regulator, computes the divergent contributions to the
symplectic form and cancels them by local counterterms intrinsic to the
boundary. The resulting renormalized symplectic structure has several
desirable properties: it yields finite generators for large
gauge transformations and diffeomorphisms, provides well-defined
Poisson brackets for asymptotic charges, and ensures integrability of
the associated conserved quantities.

%%%%%%%%%%%%%%%%%%%%%%%%%%%%%%%%%%%%%%%%%%%%%%%%%%%%%%%%%%%%%%%%%%%%%%%%%%%%%%%%%%%%%%%%%%%%%%%%%%%%%%%%%%%%%%%%%%%%%%%%%%%%%%%%%%%%%%%%%%%%%%%%%%%%%%%%%%%%%%%%%%%%%%%%%%%%%%%%%%%%%%%%%%%%%%%%%%%%%%%%%%%%%%%%%%%%%%%%%%%%%%%%%%%%%%%%%%%%%%%%%%%%%%%%%%%%%%%%%%%%%%%%%%%%

\section{$p$-forms}\label{sec3}
    The Lagrangian of a $p$-form gauge field $ \cB_{\m_{1}\dots\m_{p}} $ is \cite{Henneaux:1986ht}
    \begin{equation} \label{lagr_pform}
    \mathcal{L}=-\frac{\sqrt{|g|}}{2(p+1)!} H_{\mu_1 \cdots \mu_{p+1}} H^{\mu_1 \cdots \mu_{p+1}}\,,
    \end{equation}
    where $H$ denotes the field strength $ H=d\cB $. In $D=p+2$,  $ \cB_{\m_{1}\dots\m_{p}} $ is dual, on-shell, to a scalar field $ \f $ through the Hodge duality
	\begin{equation}\label{scalar_duality}
		H = \star d\f,
	\end{equation}
	 with $\star$ the Hodge star operator. By assuming the standard behaviour of a scalar field approaching null infinity
	\begin{equation}
		\phi = \cO\left({r^{-\tfrac{D-2}{2}}}\right)=\cO\left({r^{-\tfrac{p}{2}}}\right)\,,
	\end{equation} 
	the duality induces the field strength falloffs, whose Bondi components are 
	\begin{equation}\label{constraints_H}
		H_{uri_1\dots i_{p-1}}=\cO\left(r^{\tfrac{p-4}{2}}\right),\qquad  H_{ri_1\dots i_{p}} = \cO\left(r^{\tfrac{p-2}{2}}\right),\qquad  H_{ui_1\dots i_{p}} = \cO\left(r^{\tfrac{p}{2}}\right)\,.
	\end{equation}
    We assume the field strength to admit a radial-power expansion. 
    Furthermore, $p$-form gauge theories admit gauge-for-gauge redundancies that can be understood by means of the Young machinery
    \begin{equation}
		\centering
		\begin{tabular}{r@{}l}
			\raisebox{-5.8ex}{$p\left\{\vphantom{\begin{array}{c}~\\[8.8ex] ~
				\end{array}}\right.$} &
			\begin{ytableau}
				~                \\
				~    \\
				\none[\vdots]                \\
				~                   \\
				~                      
			\end{ytableau}\\[-1ex]
		\end{tabular}
			\quad \to \;\;
        \begin{tabular}{r@{}l}
			\raisebox{-4ex}{$p-1\left\{\vphantom{\begin{array}{c}~\\[8ex] ~
				\end{array}}\right.$} &
			\begin{ytableau}
				~                \\
				~    \\
				\none[\vdots]                \\
				~                   \\
				~    \pr                  
			\end{ytableau}\\[-1ex]
			
		\end{tabular}
			\quad \to \;\;
            \begin{tabular}{r@{}l}
			\raisebox{-4ex}{$p-2\left\{\vphantom{\begin{array}{c}~\\[8ex] ~
				\end{array}}\right.$} &
			\begin{ytableau}
				~                \\
				~    \\
				\none[\vdots]                \\
				~                   \\
				~    \pr                  
			\end{ytableau}\\[-1ex]
			
		\end{tabular}
        \quad \cdots\,,
	\end{equation}\\
where we start with a $p$-form gauge field Young tableaux, while the gauge transformation is represented as the derivative of an $p-1$ indexes tensor, which play the role of the gauge parameter, then completely antisymmetrised. The gauge parameter has a gauge redundancy itself, called first gauge for gauge. This chain of gauges for gauge redundancies goes on until arriving at a scalar. 

Gauge transformations are parametrised as 
\begin{equation}
    \d_\e B_{\mu_1 \dots \m_p} = \pr_{[\mu_1} \e_{\m_2 \dots \m_p ]}\,,
\end{equation}
while gauge-for-gauge parameters are always denoted with the same letter $\e$, as they are distinguished by the number of indices.  

We chose the Lorenz gauge condition for the gauge field, the gauge parameter and all the gauge for gauge parameters (except for the scalar)
    \begin{equation}
		\nabla^{\n}\cB_{\n \m_1...\m_{p-1}}=0 ,\quad \nabla^{\n}\e_{\n \m_1...\m_{p-2}}=0 ,\;\;\dots\,, \quad \nabla^{\n}\e_{\n }=0 \,,
	\end{equation}
    so that all the quantities satisfy the wave equation
     \begin{equation}
		\Box \cB_{\m_1...\m_{p}}=0 ,\quad \Box\e_{\m_1...\m_{p-1}}=0 , \;\;\cdots\,, \quad  \Box\e=0 \,.
	\end{equation}
	
    %%%%%%%%%%%%%%%%%%%%%%%%%%%%%%%%%%%%%%%%%%%%%%%%%%%%%%%%%%%%%%%%%%%%%%%%%%%%%%%%%%%%%%%%%%%%%%%%%%%%%%%%%%%%%%%%%%%%%%%%%%%%%%

	\subsection{Falloffs}
    The ansatz on the field strength expansion implies  the following behaviour of the $p$-form gauge field
	\begin{equation}\label{exp}
    \cB_{\m_1...\m_p} = B_{\m_1...\m_p} + \pr_{[\m_1}b_{\m_2...\m_p]} =\ \sum_{n,n_1,n_2}\left[\frac{B_{\m_1...\m_p}^{(n)}}{r^n}+\pr_{[\m_1}\left(b^{(n_1)}_{\m_2...\m_p]}\frac{1}{r^{n_1}}+\hat{b}^{(n_2)}_{\m_2...\m_p]}\frac{\ln{(r)}}{r^{n_2}}\right)\right] \, ,
	\end{equation}
	where the pure gauge part (exact part of the form) has been factored out, so that only the first series is relevant to
	the field strength. 
    Field strength falloffs \eqref{constraints_H}, together with the above expansion ansatz, the equations of motion and the gauge condition, determine the falloffs of $B_{\m_1...\m_p}$ to be
	\begin{equation}\label{fallB}
		\begin{split} 
			&B_{i_1\dots i_p} = \cO\left(r^{\tfrac{p}{2}}\right)\,, \;\;\;\qquad\qquad  B_{ri_1\dots i_{p-1}}  = \cO\left(r^{\tfrac{p-4}{2}}\right)\,, \\
			&B_{ui_1\dots i_{p-1}}  = \cO\left(r^{\tfrac{p-2}{2}}\right)\,,\qquad  B_{uri_1\dots i_{p-2}}  = \cO\left(r^{\tfrac{p-6}{2}}\right)\,,
		\end{split}
	\end{equation}
    with the extra conditions
    \begin{equation}
       D^i B_{ur i i_1 \dots i_{p-3}}^{\big(\tfrac{4-p}{2}\big)}=0\,.
    \end{equation}
	These are the standard radiation falloffs for a $p$-form in $D=p+2$ in Lorenz gauge.  In particular, $B_{i_1\dots i_p}^{(-\tfrac{p}{2})}$ has the role of the leading free data and we assume that it approaches a well-defined function in the $u \rightarrow -\infty$ limit and admits a $u$-power expansion near $\mathcal{I}^+_-$
    \begin{equation}\label{freeexp}
        B_{i_1\dots i_p}^{(-\tfrac{p}{2})}(u,\{x^i\})=\sum_{m=-\infty}^0u^m B_{i_1\dots i_p}^{(-\tfrac{p}{2},m)}(\{x^i\})\, .
    \end{equation}
    We note that with the radiation falloffs \eqref{constraints_H} and \eqref{fallB}, the bulk action associated to \eqref{lagr_pform} defines a well-posed variational principle at null infinity. Indeed, the boundary term in the on-shell variation can scale at most as $O(r^{-1})$ for $r\to\infty$. Therefore, the boundary contribution vanishes asymptotically and no additional boundary term is required to render the variational problem well defined\footnote{More generally if we assume falloffs \eqref{constraints_H} and \eqref{fallB} without fixing the dimension $D$, we get that the boundary terms in the variation of the action diverges if $D > p+3$.}.
    
    Moreover we consider $b$ terms admitting a polyhomogenoues expansion of the form
    \begin{equation}\label{fallb}
		\begin{aligned}
			&b_{uri_1...i_{p-3}}=\sum_{n=-N-(\frac{p-4}{2})}^{\infty}b^{(n)}_{uri_1...i_{p-3}}\frac{1}{r^n}+\sum_{n=\frac{6-p}{2}}^{\infty}\hat{b}^{(n)}_{uri_1...i_{p-3}}\frac{\ln{(r)}}{r^n} \, ,\\
			&b_{ui_1...i_{p-2}}=\sum_{n=-N-(\frac{p-2}{2})}^{\infty}b^{(n)}_{ui_1...i_{p-2}}\frac{1}{r^n}+\sum_{n=\frac{4-p}{2}}^{\infty}\hat{b}^{(n)}_{ui_1...i_{p-2}}\frac{\ln{(r)}}{r^n} \, ,\\
			&b_{ri_1...i_{p-2}}=\sum_{n=-N-(\frac{p-2}{2})}^{\infty}b^{(n)}_{ri_1...i_{p-2}}\frac{1}{r^n}+\sum_{n=\frac{4-p}{2}}^{\infty}\hat{b}^{(n)}_{ri_1...i_{p-2}}\frac{\ln{(r)}}{r^n} \, ,\\
			&b_{i_1...i_{p-1}}=\sum_{n=-N-(\frac{p}{2})}^{\infty}b^{(n)}_{i_1...i_{p-1}}\frac{1}{r^n}+\sum_{n=\frac{2-p}{2}}^{\infty}\hat{b}^{(n)}_{i_1...i_{p-1}}\frac{\ln{(r)}}{r^n} \, .\\
		\end{aligned}    
	\end{equation}
    From constraints \eqref{constraints_H}, expansions \eqref{freeexp} and the equations of motion, we have that the
    field strength approaches $\mathcal{I}^+_-$ as
    \begin{equation}\label{expHur}
        H_{uri_1...i_{p-1}}=\sum_{n=\frac{p-4}{2}}^{\infty}\frac{1}{r^n}\left[\sum_{m=0}^{n-1}u^m H_{uri_1...i_{p-1}}^{(n,m)}+h^{(n)}_{uri_1...i_{p-1}}(u,\{x^i\})\right] \, ,
    \end{equation}
    with $h^{(n)}_{uri_1...i_{p-1}}(u,\{x^i\})\rightarrow 0$ for $u \rightarrow -\infty$.

	To discuss asymptotic symmetries, we apply the splitting due to the Hodge decomposition to the  gauge parameter $\e_{i_1...i_{p-1}}$
	\begin{equation}
		\e_{i_1\dots i_{p-1}} = D_{[i_1} \e''_{i_2\dots i_{p-1}]} + \ve_{i_1\dots i_{p}}D^{i_p} \ep \, ,
		\label{hodgedecpar}
	\end{equation}
	with expansion coefficients $\e''^{(n)}_{i_2\dots i_{p-1}}, \hat{\e}''^{(n)}_{i_2\dots i_{p-1}}$ and $\e'^{(n)}, \hat{\e}'^{(n)}$. The gauge-for-gauge redundancy, $\e_{i_1...i_{p-1}} \mapsto \e_{i_1...i_{p-1}} + D_{[i_1}\e_{i_1...i_{p-2}]}$ affects only the exact part; indeed using the Hodge decomposition 
    \begin{equation}
    D_{[i_1} \e''_{i_2\dots i_{p-1}]} + \ve_{i_1\dots i_{p}}D^{i_p} \e \mapsto  D_{[i_1} \e''_{i_2\dots i_{p-1}]} + \ve_{i_1\dots i_{p}}D^{i_p} \e + D_{[i_1}\e_{i_1...i_{p-2}]} \, ,
    \end{equation}
    and using that the decomposition is orthogonal we get that $\e'$ is unvariant while $ \e''_{i_2\dots i_{p-1}} \rightarrow  \e''_{i_2\dots i_{p-1}}+  \e_{i_2\dots i_{p-1}}$.
    Similar expansions hold for the components
	\begin{equation}
		\begin{aligned}
			B_{ui_1...i_{p-1}}&=D_{[i_1} B''_{ui_2\dots i_{p-1}]} + \ve_{i_1\dots i_{p}}D^{i_p} B^\p_u \, , \\
			B_{ri_1...i_{p-1}}&=D_{[i_1} B''_{ri_2\dots i_{p-1}]} + \ve_{i_1\dots i_{p}}D^{i_p} B^\p_r \, ,
		\end{aligned}    
	\end{equation}
	while
	\begin{equation}
		B_{i_1...i_{p}}=\ve_{i_1...i_{p}}B \, .
	\end{equation}
	Performing a gauge transformation parametrised by $\e''_{i_2\dots i_{p-1}}$ and $\ep$ we get
	\begin{equation}
		\delta_{\e''_{i_2\dots i_{p-1}}}B=0, \qquad \delta_{\ep}B=\Delta \ep \, .
	\end{equation}

%%%%%%%%%%%%%%%%%%%%%%%%%%%%%%%%%%%%%%%%%%%%%%%%%%%%%%%%%%%%%%%%%%%%%%%%%%%%%%%%%%%%%%%%%%%%%%%%%%%%%%%%%%%%%%%%%%%%%%%%%%%%%%%%%%%%%%%%%%%%%%%%%%%%%%%%%%%%%%%%%%%%%%%%%%%%%%%%%%%%%%%%%%%%%%%%%%%%%%%%%

\subsection{Asymptotic symmetries and charges}
	Considering a constant time surface $ \S_t $, we can define a charge $Q_t$ as
    \begin{equation}
	Q_{t}:= \int_{\S_t}du d\O_{D-2} \; \cJ^t 
	\end{equation}
    with $\cJ$ being the Noether current
    \begin{equation}
        \cJ^t = \frac{1}{(p-1)!}  \pr_\m\left (\sqrt{|g|} H^{t\m \m_1\dots \m_{p-1}}\e_{\m_1 \dots \m_{p-1}}\right)\,.
    \end{equation}
    Neglecting a total $(D-2)$-sphere divergence, we can write $Q_{t}$ respect to the Bondi components, namely
    \begin{equation}\label{B_charge}
	Q_{t}:=\frac{1}{(p-1)!} \int_{ \S_t} du d\O_{D-2} \; (\pr_u -\pr_r)\left (r^{D-2} H^{uri_1 \dots i_{p-1}}\e_{i_1 \dots i_{p-1}}\right)\,.
    \end{equation}
    The asymptotic charge is thus defined taking the limit $t\to\infty$ at fixed $u$, so that $\S_t$ approaches $\cI^+$, this is equivalent to the limit $r \rightarrow \infty$. However, for higher-order asymptotic symmetries there are divergences which make the limit ill-defined. Symplectic renormalization is employed to cancel these divergent terms and perform the limit. The asymptotic charge is thus defined as
	\begin{equation}\label{as_charge}
	Q_{B}:=\lim_{t\to\infty} Q_{t, \text{ren}}\,,
	\end{equation}
    where we denoted with $Q_{t, \text{ren}}$ the symplectic-renormalized version (details in \autoref{sec4}) of the charge \eqref{B_charge} and the limit is taken at constant $u$.
Given the Hodge decomposition of the gauge parameter \eqref{hodgedecpar}
	\begin{equation}\label{Hodge_par}
		\begin{split}
			\e_{i_1\dots i_{p-1}} =& D_{[i_1} \e^{\p\p}_{i_2\dots i_{p-1}]}+ \ve_{i_1\dots i_{p}} D^{i_p} \e^\p\,,
		\end{split}
	\end{equation}
    and of the field strength component
    \begin{equation}\label{hodgeex}
			H_{uri_1\dots i_{p-1}} = D_{[i_1} H^{\p\p}_{i_2\dots i_{p-1}]ur}+ \ve_{i_1\dots i_{p}} D^{i_p} H^\p_{ur}\,,
	\end{equation}
	the charge reads simply
	\begin{equation}
		Q_t = -\int_{\S_t}du d\O_{D-2} \; (\pr_u -\pr_r)\left ( r^{D-2}\e^\p \D H_{ur}^\p\right)\,,
	\end{equation}
    where we used
	\begin{equation}
		D^{i_1} H_{ur i_1\dots i_{p-1} } =0\,,
	\end{equation}
	which originates from the Bianchi identity $ \nabla_{[i_1} H_{ur i_2\dots i_{p}] } =0$. 

Therefore, the only relevant component is $\e^\p$. To study $\cO(r^N)$ asymptotic symmetries, we consider this expansion
\begin{equation}
    \e^\p=\sum_{n=-N-(\frac{p}{2})}^{\infty}\frac{\e^{\p(n)}}{r^n}+\sum_{n=\frac{2-p}{2}}^{\infty}\hat{\e}^{\p(n)}\frac{\ln{(r)}}{r^n} \, .
\end{equation}
The logarithmic series, which is subleading, is required to find an order $\e^{\p(-p/2)}$ with an arbitrary angular dependence. In fact, by studying the equations of motion \eqref{gauge_par_eom} for $\e^\p$ at that specific order, we obtain
\begin{equation}
    2\pr_u\he^{\p \,\big(\tfrac{2-p}{2}\big)}=\left[ \D -\frac{p(p-2)}{4} \right]\e^{\p\,\big(-\tfrac{p}{2}\big)}\,.
\end{equation}
If we do not consider the logarithmic series, namely $\he^{\p \,\big(\tfrac{2-p}{2}\big)}=0$, then $\e^{\p\,\big(-\tfrac{p}{2}\big)}$ is constrained by the previous equation. Focusing only on the overleading part with $-(N+p/2)\leq n \leq -p/2$, which is the one that gives a contribution to the asymptotic charge, the wave equation gives
\begin{equation}\label{u_structure_par}
    \e^{\p (-n-p/2)}(u,\{x^i\}) = \sum_{m=0}^{N-n} u^m \e^{\p (-n-p/2,m)}(\{x^i\})\,,
\end{equation}
where we remark that this solution is valid for all values of $u$. The only free functions are the $N+1$ $u$-integration functions $\e^{\p (-n-p/2,0)}(\{x^i\})$, which completely determine the structure of all the  $\e^{\p (-n-p/2,m)}$ with $m>1$.

These integration functions parametrise the asymptotic charges. In fact, the renormalization procedure, discussed in details in \autoref{sec4}, leaves us with $N+1$ asymptotic charges
\begin{equation}
    Q_B = \sum_{k=0}^N Q_B^{(k)}\,,
\end{equation}
with each charge being
\begin{equation}
    Q_B^{(k)} =  \int_{\cI^+_-} d\O_{D-2} \e^{\p \big(-\tfrac{p}{2}-k,0\big)} \D H_{ur}^{\p \big(\tfrac{4-p}{2}+k,0\big)}\,.
\end{equation}
The integration over ${\cI^+_-}$ deserves some comments. Although the limiting procedure on $\Sigma_t$ will produce an integral over space-like infinity, we assume, as often in literature, that charges computed at ${\cI^+_-}$ correspond to those at space-like infinity. Some papers that study this issue more
in detail in gravity and in gauge theory are, for example \cite{Ashtekar:1978zz,Campiglia:2015qka,Campiglia_2017}.Another point is to clarify the role of falloffs at different values of $u$. In general, the $u$-integral is divergent: the assumed $u$-falloffs behaviour of the fields is therefore essential to control and to identify the divergent contributions as well as to define the appropriate renormalization counterterms in the next \autoref{sec4}. 

Performing a small gauge fixing, a proper gauge transformation that is
associated to vanishing charges, at the boundary, as detailed in Appendix \ref{AppB}, we are able to determine the expression in terms of the Cauchy data. Specifically, we find for even $p$
 \begin{equation}
       \begin{split}
            H_{ur}^{\p\big(\tfrac{4-p}{2},0\big)} &= - B^{(-{p}/{2},0)} \,,\\
             H_{ur}^{\p(2,0)} \quad\;&= \frac{\D}{p} B_r^{\p(1,0)}\,,\\
           H_{ur}^{\p(n,0)} \quad\; &=  \frac{n-2}{n+p-2} B^{(n-2,0)} \qquad n\neq \tfrac{4-p}{2},\, 2\;\;,
       \end{split} 
    \end{equation}
    where the component $H_{ur}^{\p(2,0)}$ can be determined only in terms of $B_r^{\p(1,0)}$, which is the natural Cauchy data at that order, since $B^{(0,0)}$ is a pure gauge term. For odd $p$, $n$ is semi-integer and we can determine everything in terms of $B$
    \begin{equation}
       \begin{split}
            H_{ur}^{\p\big(\tfrac{4-p}{2},0\big)} &= +B^{(-{p}/{2},0)} \,,\\
           H_{ur}^{\p(n,0)} \quad\; &=  \frac{n}{n+p-2} B^{(n-2,0)} \qquad n\neq \tfrac{4-p}{2} \;\;.
       \end{split} 
    \end{equation}
    \subsection{Explicit examples}
    In this paragraph we make explicit some cases.
	\subsubsection*{1-form in $D=3$ with $N=4$}
    We start with a maxwell field $p=1$, so that $D = p + 2 = 3$; the celestial sphere is \(S^{1}\). In addition, taking \(N = 4\) produces a tower of \(N+1 = 5\) renormalized charges. For \(p = 1\) the gauge field is a \(1\)-form \(B_{\mu}\) with field strength
\begin{equation}
H_{\mu\nu} = \partial_{\mu} B_{\nu} - \partial_{\nu} B_{\mu} \, .
\label{eq:H_def_p1}
\end{equation}
The radiative falloffs induced by duality with a scalar become, in Bondi coordinates \((u,r,\varphi)\),
\begin{equation}
H_{ur} = O\!\left(r^{-3/2}\right) \, ,\qquad
H_{r\varphi} = O\!\left(r^{-1/2}\right) \, ,\qquad
H_{u\varphi} = O\!\left(r^{1/2}\right) \, ,
\label{eq:falloffs_H_p1}
\end{equation}
and for the potential one has
\begin{equation}
B_{\varphi} = O\!\left(r^{1/2}\right) \, ,\qquad
B_{u} = O\!\left(r^{-1/2}\right) \, ,\qquad
B_{r} = O\!\left(r^{-3/2}\right) \, .
\label{eq:falloffs_B_p1}
\end{equation}

Here the Hodge decomposition on the sphere becomes trivial for the gauge parameter, because for \(p = 1\) the parameter \(\epsilon\) is already a scalar on \(S^{1}\) . In particular, the sector usually denoted \(\epsilon'\) in the general discussion coincides with \(\epsilon\) when \(p = 1\).

With overleading order \(N = 4\) , the relevant expansion of the gauge parameter contains five independent coefficient functions (integration functions) ,
\begin{equation}
\epsilon^{\p \big(-\tfrac12-k,\,0\big)}(\varphi) \qquad k = 0,1,2,3,4 \, ,
\label{eq:epsilon_coeffs_N4}
\end{equation}
and these parametrize the five renormalized charges. The Laplacian is simply
\begin{equation}
\Delta = \partial_{\varphi}^{2} \, ;
\label{eq:laplacian_S1}
\end{equation}
therefore, we have
\begin{equation}
Q^{(k)}_B
=
\int_{\cI^+_-} d\varphi \;
\epsilon^{\p \big(-\tfrac12-k,\,0\big)}(\varphi)\;
\partial_{\varphi}^2\, H_{ur}^{\p \big(\tfrac32+k,\,0\big)}(\varphi)
\qquad
k = 0,\dots,4 \, .
\label{eq:Qk_in_terms_of_Hur}
\end{equation}

Finally, because \(p = 1\) is odd, the coefficients \(H_{ur}^{(n,0)}\) can be rewritten directly in terms of the Cauchy data (the scalar mode associated to the angular component). Concretely,
\begin{equation}
H_{ur}^{\p \big(\tfrac32,\,0\big)} = B^{\big(-\tfrac12,\,0\big)} \, ,
\label{eq:Hur32_equals_Bminus12}
\end{equation}
and for \(k = 1,2,3,4\) we have
\begin{equation}
H_{ur}^{\p \big(\tfrac32+k,\,0\big)}
=
\frac{\tfrac32+k}{\tfrac12+k}\, B^{\big(-\tfrac12+k,\,0\big)}
=
\frac{2k+3}{2k+1}\, B^{\big(-\tfrac12+k,\,0\big)} \, .
\label{eq:Hur_coeffs_in_terms_of_B}
\end{equation}

Using \eqref{eq:Hur32_equals_Bminus12} and \eqref{eq:Hur_coeffs_in_terms_of_B} in \eqref{eq:Qk_in_terms_of_Hur} gives the five charges explicitly

\begin{equation}
Q^{(0)}_B
=
\int_{\cI^+_-} d\varphi \;
\epsilon^{\p(-\tfrac12,\,0)}(\varphi)\;
\partial_\varphi^{2} B^{(-\tfrac12,\,0)}(\varphi) \, .
\label{eq:Q0}
\end{equation}

\begin{equation}
Q^{(1)}_B
=
\int_{\cI^+_-} d\varphi \;
\epsilon^{\p(-\tfrac32,\,0)}(\varphi)\;
\partial_\varphi^{2}
\!\left(
\frac{5}{3}\, B^{(\tfrac12,\,0)}(\varphi)
\right) \, .
\label{eq:Q1}
\end{equation}

\begin{equation}
Q^{(2)}_B
=
\int_{\cI^+_-} d\varphi \;
\epsilon^{\p(-\tfrac52,\,0)}(\varphi)\;
\partial_\varphi^{2}
\!\left(
\frac{7}{5}\, B^{(\tfrac32,\,0)}(\varphi)
\right) \, .
\label{eq:Q2}
\end{equation}

\begin{equation}
Q^{(3)}_B
=
\int_{\cI^+_-} d\varphi \;
\epsilon^{\p(-\tfrac72,\,0)}(\varphi)\;
\partial_\varphi^{2}
\!\left(
\frac{9}{7}\, B^{(\tfrac52,\,0)}(\varphi)
\right) \, .
\label{eq:Q3}
\end{equation}

\begin{equation}
Q^{(4)}_B
=
\int_{\cI^+_-} d\varphi \;
\epsilon^{\p(-\tfrac92,\,0)}(\varphi)\;
\partial_\varphi^{2}
\!\left(
\frac{11}{9}\, B^{(\tfrac72,\,0)}(\varphi)
\right) \, .
\label{eq:Q4}
\end{equation}
Equivalently, since the prefactors are constants, one may pull them out and write the compact form
\begin{equation}
Q^{(k)}_B
=
c_k
\int_{\cI^+_-} d\varphi \;
\epsilon^{\p(-\tfrac12-k,\,0)}(\varphi)\;
\partial_\varphi^{2} B^{(-\tfrac12+k,\,0)}(\varphi) \, ,
\label{eq:Qk_compact}
\end{equation}
with
\begin{equation}
c_0 = 1 \, ,\qquad
c_1 = \frac{5}{3} \, ,\qquad
c_2 = \frac{7}{5} \, ,\qquad
c_3 = \frac{9}{7} \, ,\qquad
c_4 = \frac{11}{9} \, .
\label{eq:ck_values}
\end{equation}
    \subsubsection*{2-form in $D=4$ with $N=3$}
    Let us consider the case of a $2$-form in $D=p+2=4$. Therefore the celestial sphere is \( S^{2}\) with angular coordinates \(x:=(\varphi,\vartheta)\); the tower of renormalized charges with $N=3$ counts \(N+1 = 4\) elements \(Q^{(k)}_B\) with \(k = 0,1,2,3\) . For \(p = 2\) the gauge field is a \(2\)-form \(B_{\mu\nu}\) with field strength
\begin{equation}
H_{\mu\nu\rho} =\partial_{[\mu} B_{\nu\rho]} \, .
\label{eq:H_def_p2}
\end{equation}
The radiative falloffs induced by duality with a scalar become, in Bondi coordinates
\begin{equation}
H_{ur i} = O\!\left(r^{-1}\right) \, ,\qquad
H_{r i j} = O\!\left(r^{0}\right) \, ,\qquad
H_{u i j} = O\!\left(r^{1}\right) \, ,
\label{eq:falloffs_H_p2}
\end{equation}
while the gauge potential components obey
\begin{equation}
B_{i j} = O\!\left(r^{1}\right) \, ,\qquad
B_{r i} = O\!\left(r^{-1}\right) \, ,\qquad
B_{u i} = O\!\left(r^{0}\right) \, ,\qquad
B_{u r} = O\!\left(r^{-2}\right) \, .
\label{eq:falloffs_B_p2}
\end{equation}
The Hodge decomposition on \(S^{2}\) implies that 
\begin{equation}
    \e_i=D_i\e''+\varepsilon_{ij}D^j\ep 
\end{equation}
and only the scalar sector $\ep$ of the gauge parameter contributes to the charge. For order \(O(r^{3})\) asymptotic symmetries, the relevant overleading part of \(\epsilon'\) contains \(N+1=4\) free coefficient functions on \(S^{2}\) ,
\begin{equation}
\epsilon^{\p(-1-k,\,0)}(x)\qquad k = 0,1,2,3 \, ,
\label{eq:epsilon_coeffs_p2N3}
\end{equation}
and these are precisely the functions that parametrize the four renormalized charges, which are given by
\begin{equation}
Q^{(k)}_B
=
\int_{\cI^+_-} d\Omega_{2}\;
\epsilon^{\p(-1-k,\,0)}(x)\;
\Delta\, H_{ur}^{\p \,(1+k,\,0)}(x)
\qquad
k = 0,1,2,3 \, ,
\label{eq:Qk_in_terms_of_Hur_p2}
\end{equation}
where \(H'_{ur}\) is the scalar obtained by Hodge-decomposing the sphere index of \(H_{ur i}\) .

For even \(p\) one can perform a small gauge fixing at the boundary and rewrite the relevant coefficients \(H_{ur}^{\p(n,0)}\) in terms of Cauchy data. In the present case \(p=2\) is even, hence
\begin{equation}
H_{ur}^{\p(1,0)} = -\,B^{(-1,0)} \, .
\label{eq:Hur10_p2}
\end{equation}
\begin{equation}
H_{ur}^{\p(2,0)} = \frac{\Delta}{2}\, B_r^{\p(1,0)} \, .
\label{eq:Hur20_p2}
\end{equation}
and for \(k = 2,3\) we have
\begin{equation}
H_{ur}^{\p(1+k,0)}
=
\frac{k-1}{k+1}\,B^{(-1+k,0)} \, .
\label{eq:Hur_n0_p2_general}
\end{equation}
Plugging \eqref{eq:Hur10_p2} , \eqref{eq:Hur20_p2} and \eqref{eq:Hur_n0_p2_general} into \eqref{eq:Qk_in_terms_of_Hur_p2} gives the four charges explicitly
\begin{equation}
Q^{(0)}_B
=
\int_{\cI^+_-} d\Omega_{2}\;
\epsilon^{\p(-1,\,0)}(x)\;
\Delta\!\left(-\,B^{(-1,0)}(x)\right) \, .
\label{eq:Q0_p2}
\end{equation}

\begin{equation}
Q^{(1)}_B
=
\int_{\cI^+_-} d\Omega_{2}\;
\epsilon^{\p(-2,\,0)}(x)\;
\Delta\!\left(\frac{\Delta}{2}\,B_{r}^{\p(1,0)}(x)\right) \, .
\label{eq:Q1_p2}
\end{equation}

\begin{equation}
Q^{(2)}_B
=
\int_{\cI^+_-} d\Omega_{2}\;
\epsilon^{\p(-3,\,0)}(x)\;
\Delta\!\left(\frac{1}{3}\,B^{(1,0)}(x)\right) \, .
\label{eq:Q2_p2}
\end{equation}

\begin{equation}
Q^{(3)}_B
=
\int_{\cI^+_-} d\Omega_{2}\;
\epsilon^{\p(-4,\,0)}(x)\;
\Delta\!\left(\frac{1}{2}\,B^{(2,0)}(x)\right) \, .
\label{eq:Q3_p2}
\end{equation}
The charge \(Q^{(1)}\) remains distinguished, because the even-\(p\) boundary gauge fixing leaves \(B_{r}^{\p(1,0)}\) as the natural Cauchy datum at that order.

In Maxwell theory in $D=4$, something similar happens: we can construct $\mathcal{O}(r^N)$ asymptotic symmetries which corresponds to a tower of asymptotic charges \cite{Romoli:2024hlc,Peraza:2023ivy}. It would be interesting understand how the infinite tower of charges $Q_B^{(k)}$ enter into soft theorems for the 2-form field.

    \subsubsection*{3-form in $D=5$ with $N=2$}
    We now focus on a 3-form, hence $D = p + 2 = 5$ and the celestial sphere is \(S^{D-2} = S^{3}\) with angular coordinates \(x:=(\varphi,\vartheta,\chi)\). Taking $N=2$ the tower of renormalized charges has \(N+1 = 3\) elements. For \(p=3\) the gauge field is a \(3\)-form \(B_{\mu\nu\rho}\) with field strength
\begin{equation}
H_{\mu\nu\rho\sigma} = \partial_{[\mu} B_{\nu\rho\sigma]} \, .
\label{eq:H_def_p3}
\end{equation}
Assuming the standard scalar radiation behaviour in and using duality, the field strength falloffs become
\begin{equation}
H_{urij} = O\!\left(r^{-1/2}\right) \, ,\qquad
H_{rijk} = O\!\left(r^{1/2}\right) \, ,\qquad
H_{uijk} = O\!\left(r^{3/2}\right) \, .
\label{eq:falloffs_H_p3}
\end{equation}

Moreover, the Lorenz gauge condition and the equations of motion determine the gauge potential falloffs as
\begin{equation}
B_{ijk} = O\!\left(r^{3/2}\right) \, ,\qquad
B_{rij} = O\!\left(r^{-1/2}\right) \, ,\qquad
B_{uij} = O\!\left(r^{1/2}\right) \, ,\qquad
B_{uri} = O\!\left(r^{-3/2}\right) \, .
\label{eq:falloffs_B_p3}
\end{equation}
The gauge parameter admits the Hodge decomposition on \(S^{3}\), which in components reads
\begin{equation}
\epsilon_{ij}
=
D_{[i}\epsilon''_{j]}
+
\varepsilon_{ijk}\,D^{k}\epsilon' \, .
\label{eq:hodge_eps_p3}
\end{equation}
Only the scalar sector \(\epsilon'\) contributes to the asymptotic charge. At order \(O(r^{2})\), the overleading part of \(\epsilon'\) is determined by the \(N+1=3\) functions
\begin{equation}
\epsilon^{\p(-\tfrac{3}{2}-k,\,0)}(x)\qquad k = 0,1,2 \, ,
\label{eq:eps_coeffs_p3N2}
\end{equation}
and these parametrize the three renormalized charges, which are given by
\begin{equation}
Q^{(k)}_B
=
\int_{\cI^+_-} d\Omega_{3}\;
\epsilon^{\p(-\tfrac{3}{2}-k,\,0)}(x)\;
\Delta\, H_{ur}^{\p(\tfrac{1}{2}+k,\,0)}(x)
\qquad
k = 0,1,2 \, .
\label{eq:Qk_universal_p3}
\end{equation}

Here \(H'_{ur}\) is the scalar entering the Hodge decomposition of \(H_{urij}\) on \(S^{3}\). For odd \(p\) the relevant coefficients can be expressed entirely in terms of the Cauchy data \(B\)
\begin{equation}
H_{ur}^{\p(\tfrac{1}{2},\,0)} = B^{(-\tfrac{3}{2},\,0)} \, ,
\label{eq:Hur_half_p3}
\end{equation}
\begin{equation}
H_{ur}^{\p(\frac{1}{2}+k,\,0)}
=
\frac{\frac{1}{2}+k}{\frac{3}{2}+k}\,B^{(-\frac{3}{2}+k,\,0)} = \frac{2k+1}{2k+3}\,B^{(-\frac{3}{2}+k,\,0)}\, .
\label{eq:Hur_n_p3_general}
\end{equation}
Plugging \eqref{eq:Hur_half_p3} and \eqref{eq:Hur_n_p3_general} into \eqref{eq:Qk_universal_p3} yields
\begin{equation}
Q^{(0)}_B
=
\int_{\cI^+_-} d\Omega_{3}\;
\epsilon^{\p(-\tfrac{3}{2},\,0)}(x)\;
\Delta\, B^{(-\tfrac{3}{2},0)}(x) \, .
\label{eq:Q0_p3}
\end{equation}
\begin{equation}
Q^{(1)}_B
=
\int_{\cI^+_-} d\Omega_{3}\;
\epsilon^{\p(-\tfrac{5}{2},\,0)}(x)\;
\Delta\!\left(\frac{3}{5}\,B^{(-\tfrac{1}{2},0)}(x)\right) \, .
\label{eq:Q1_p3}
\end{equation}

\begin{equation}
Q^{(2)}_B
=
\int_{\cI^+_-} d\Omega_{3}\;
\epsilon^{\p(-\tfrac{7}{2},\,0)}(x)\;
\Delta\!\left(\frac{5}{7}\,B^{(\tfrac{1}{2},0)}(x)\right) \, .
\label{eq:Q2_p3}
\end{equation}
	\section{Symplectic renormalization of the charge}\label{sec4}
	The study of $\cO(r^N)$ asymptotic symmetries and related charges naturally leads to divergences that are of two types: those arising from
	positive powers of $t$ and those resulting from the $du$ integration. To address these issues, we employ a symplectic renormalization procedure, as proposed in \cite{Freidel:2019ohg}. \\
	As we discussed in the preliminaries, the presymplectic potential admits two types of ambiguities that allow us to define an
	equivalence relation 
	\begin{equation}
		\theta^{\m}\sim_{\text{eq}}\theta^{\m}+\pr_{\n}\Upsilon^{\m \n}+\delta\Xi^{\m} \, ,
	\end{equation}
	with $\Upsilon^{\m \n}$ an antisymmetric rank-two tensor. Our main goal is to use these ambiguities to cancel the
	divergences in the asymptotic charges while leaving the rest unchanged. \\
	In $(u,t,\{x^i\})$ coordinates we have 
	\begin{equation}
		\pr_t\theta^t \approx \delta \mathcal{L}-\pr_u\theta^u-\pr_i\theta^i \, ,
        \label{ambpresym}
	\end{equation}
	where ``$\approx$'' denotes an on-shell relation. We can show that in the limit $t \rightarrow +\infty $ we have that $\theta^t$ admits, as already observed in \cite{Romoli:2024hlc} for the $p=2$ case in $D=4$, a general structure. Hence we have the following general structure for $\theta^t$
        \begin{equation}\label{exp1}
            \theta^t=\theta^t_{(0)}(u,t,\{x^i\})+\sum_{n=1}^N t^n\theta^t_{(n)}(u,\{x^i\}) \, ,
        \end{equation}
        with $\theta^t_{(0)}(u,t,\{x^i\})$ such that $\lim_{t \rightarrow +\infty}\theta^t_{(0)}(u,t,\{x^i\})=\theta^t_{(0)}(u,\{x^i\})$.
    Indeed, the presymplectic potential is given by
	\begin{equation}\label{thetaexp}
		\begin{aligned}
			\theta^t&=\theta^u+\theta^r=r^{D-2}\frac{|\gamma|}{p!}(H^{u\m_1...\m_p}+H^{r\m_1...\m_p})\delta \cB_{\m_1...\m_p}
            \\
			&=\frac{|\gamma|}{r^p}\left[r^2{\prod_{k=1}^{p-1}\gamma^{i_kj_k}}H_{uri_1...i_{p-1}}(\delta \cB_{uj_1...j_{p-1}}-\delta \cB_{rj_1...j_{p-1}})-{\prod_{k=1}^p\gamma^{i_kj_k}}H_{ui_1...i_p}\delta \cB_{j_1...j_p}\right] \, ,
		\end{aligned}
	\end{equation}
	where $|\gamma|:=\text{det}(\gamma_{ij})$.	The field strength components admit a power expansion with falloffs \eqref{constraints_H}
    	\begin{equation}
		H_{uri_1\dots i_{p-1}}=\cO\left(r^{\tfrac{p-4}{2}}\right),\qquad  H_{ri_1\dots i_{p}} = \cO\left(r^{\tfrac{p-2}{2}}\right),\qquad  H_{ui_1\dots i_{p}} = \cO\left(r^{\tfrac{p}{2}}\right)\,.
	\end{equation}
    Moreover, the variation of the $p$-form field is
	\begin{equation}
		\delta \cB_{\m_1...\m_p}=\delta B_{\m_1...\m_p}-\pr_{[\m_1}\delta b_{\m_2...\m_p]}
	\end{equation}
	with 
    \begin{equation}
		\begin{split} 
			&B_{i_1\dots i_p} = \cO\left(r^{\tfrac{p}{2}}\right)\,, \;\;\;\qquad\qquad  B_{ri_1\dots i_{p-1}}  = \cO\left(r^{\tfrac{p-4}{2}}\right)\,, \\
			&B_{ui_1\dots i_{p-1}}  = \cO\left(r^{\tfrac{p-2}{2}}\right)\,,\qquad  B_{uri_1\dots i_{p-2}}  = \cO\left(r^{\tfrac{p-6}{2}}\right)\,,
		\end{split}
	\end{equation}
    with a power expansion, and $b$ terms admitting a polyhomogenoues expansion of the form \eqref{fallb}. Inserting them into \eqref{thetaexp} we have three kind of contributions. The first one is given, schematically, by 
\begin{equation}\label{pres_1}
    \begin{split}
    &\frac{1}{r^{p}}H_{ui_1...i_p}\delta B_{j_1...j_p} =\cO(r^0) \, ,\\
 &\frac{1}{r^{p-2}}H_{uri_1...i_{p-1}}\delta B_{uj_1...j_{p-1}} =\cO(r^{-1}) \, ,\\
      & \frac{1}{r^{p-2}} H_{uri_1...i_{p-1}}\delta B_{rj_1...j_{p-1}} =\cO(r^{-2}) \,.
    \end{split}
\end{equation}
The second kind of terms are, schematically
\begin{equation}\label{pres_2}
   \begin{split}
       &\frac{ 1}{r^{p}}H_{ui_1...i_p}\partial_{[j_1}\delta b_{j_2...j_{p}]}= \cO(r^N) \, ,\\
       &\frac{1}{r^{p-2}}H_{uri_1...i_{p-1}}\pr_{[u} b_{j_1...j_{p-1}]} = \cO(r^{N-1}) \, ,\\
       & \frac{1}{r^{p-2}}H_{uri_1...i_{p-1}}\pr_{[r}b_{j_1...j_{p-1}]} =\cO(r^{N-1}) \,.
   \end{split}
\end{equation}
The last kind of terms are those containing logarithms which are, schematically 
\begin{equation}\label{pres_3}
   \begin{split}
       &\frac{1}{r^{p}}H_{ui_1...i_p}\partial_{[j_1}\delta \hb_{j_2...j_{p}]}= \cO(\tfrac{\ln(r)}{r}) \, ,\\
       &\frac{1}{r^{p-2}}H_{uri_1...i_{p-1}}\pr_{[u} \hb_{j_1...j_{p-1}]} = \cO(\tfrac{\ln(r)}{r}) \, ,\\
       &\frac{1}{r^{p-2}}H_{uri_1...i_{p-1}}\pr_{[r}\hb_{j_1...j_{p-1}]} =\cO(\tfrac{\ln(r)}{r^2}) \,.
   \end{split}
\end{equation}
Hence, logarithmic terms are vanishing in the $t \rightarrow +\infty$ limit and can be collected into $\theta^t_{(0)}(u,t,\{x^i\})$ together with the $\cO(1)$ terms and other possible vanishing terms. In this way we have 
\begin{equation}
    \lim_{t \rightarrow +\infty}\theta^t_{(0)}(u,t,\{x^i\})=\theta^t_{(0)}(u,\{x^i\})\, .
\end{equation}
    The term $\theta^t_{(0)}(u,\{x^i\})$ receives contributions from the first of \eqref{pres_1} and from \eqref{pres_2}. In the case $N=0$, to which we refer as ``ordinary'' asymptotic symmetries, we have simply
\begin{equation}\label{theta0}
    \theta^t_{(0)}(u,\{x^i\}):=|\gamma|{\prod_{k=1}^p\gamma^{i_kj_k}}\partial_u B^{(-\tfrac{p}{2})}_{i_1...i_p}\delta \cB^{(-\tfrac{p}{2})}_{j_1...j_p} \qquad \text{ for } N=0.
\end{equation}

The equation \eqref{ambpresym}, implies that $\theta_{(n)}^t(u,\{x^i\})$ with $n > 0$ is an ambiguity that is given by
\begin{equation}
    \theta_{(n)}^t(u,\{x^i\})=\frac{1}{n}\left(\delta \mathcal{L}_{(n-1)}-\pr_u\theta^u_{(n-1)}-\pr_i\theta^i_{(n-1)} \right) \, .
\end{equation}
From the above equation, we can identify
\begin{equation}
\begin{aligned}
    &\Upsilon^{tu}=\sum_{n=1}^N\frac{t^n}{n}\theta_{(n-1)}^u+\Upsilon^{tu}_{(0)}\, , \\
    &\Upsilon^{ti}=\sum_{n=1}^N\frac{t^n}{n}\theta_{(n-1)}^i+\Upsilon^{ti}_{(0)}\, , \\
    &\Xi^{t}=\sum_{n=1}^N-\frac{t^n}{n}\mathcal{L}_{(n-1)}+\Xi^{t}_{(0)}\, , \\
\end{aligned}
\end{equation}
where $\Upsilon^{tu}_{(0)}, \Upsilon^{ti}_{(0)}, \Xi^{t}_{(0)}$ are free functions with a well-defined behaviour in the $t \rightarrow +\infty$ limit. Therefore we have
\begin{equation}
    \theta^t \sim_{\text{eq}} \theta^t+\partial_u\Upsilon^{tu}_{(0)}+ \partial_i\Upsilon^{ti}_{(0)}+\delta \Xi^{t}_{(0)} \, ,
\end{equation}
which means that only the $t^0$ finite order contributes to the charge in the $t \rightarrow +\infty$ limit. 

The next issue is to address the $u$-divergences that arise from the $du$ integration. These divergences are cured using $\Upsilon^{t\mu}_{(0)}$. As first step we can show that $\theta^t_{(0)}$ admits the general structure 
    \begin{equation}\label{exp2}
\theta^t_{(0)}=\sum_{n=1}^N\theta_{(0,n)}^tu^{n-2}+\Theta_{0}^t(u,\{x^i\})\, ,
 \end{equation}
    where $\Theta_{0}^t(u,\{x^i\})$ is the only $u$-integrable part of $\theta^t_{(0)}$. It is enough to show the statement for the $N=0$ case since if $N>0$ we are just adding positive powers of $u$ to the field expansions. From \eqref{theta0} we have 
    \begin{equation}
        \theta^t_{(0)}(u,\{x^i\}):=|\gamma|{\prod_{k=1}^p\gamma^{i_kj_k}}\partial_u B^{(-\tfrac{p}{2})}_{i_1...i_p}\delta B^{(-\tfrac{p}{2})}_{j_1...j_p},
    \end{equation}
    while from \eqref{freeexp} we have 
    \begin{equation}
    \begin{aligned}
        &B^{(-\tfrac{p}{2})}_{j_1...j_p}=B^{(-\tfrac{p}{2},0)}_{j_1...j_p}+\cO(u^{-1})\, ,\\
        &\partial_uB^{(-\tfrac{p}{2})}_{i_1...i_p}= \cO(u^{-2}) \, .
     \end{aligned}    
    \end{equation}
    Therefore 
    \begin{equation}
        \theta^t_{(0)}(u,\{x^i\})=\cO(u^{-2}).
    \end{equation}
    
We now use $\Upsilon^{t\mu}_{(0)}$ to cancel the first series of $\theta_{(0)}^t$. To do so we can employ, for instance, these expansions
\begin{equation}
    \Upsilon^{tu}_{(0)}=\sum_{n=3}^{N+1}u^{n-2}\Upsilon^{tu}_{(0,n)}, \qquad \Upsilon^{ti}_{(0)}=\frac{1}{u}\Upsilon^{ti}_{(0,-1)}
\end{equation}
where
\begin{equation}
    \Upsilon^{tu}_{(0,n)}=-\frac{\theta^t_{(0,n+1)}}{n-2}, \qquad \partial_i\Upsilon^{ti}_{(0,-1)}=-\theta^t_{0,1}.
\end{equation}
Here we are assuming that $\Upsilon^{tu}_{(0)}$ has a $u$-power expansion, therefore it is impossible to cancel a term of the type $\frac{1}{u}$ with a $u$-derivative. To our understanding, this choice is not unique and the renormalization procedure depends on the specific structure of the renormalizing terms. We underline that in showing \eqref{exp2} the fundamental assumption is the expansion \eqref{freeexp} for the free Cauchy data at order $\cO\left(r^{\tfrac{p}{2}}\right)$. This is a power-law Laurent series which does not contain logarithmic terms. Such an expansion can also be considered and would lead to some differences in the renormalization procedure that we can investigate in the future.

In the end, this process leaves us with well-defined quantities, thus showing that it is possible to renormalize the presymplectic potential to make the asymptotic charge finite while accommodating arbitrary powers of the asymptotic parameter. Reminding equation \eqref{Noether-two-form} we are able to write the asymptotic charge as an integral over $\cI^+_-$ and, specifically, it is given by the $t-$ and  $u-$independent parts of the Noether two-form component $\k^{tu}$
\begin{equation}
    Q_B = \sum_{k=0}^N \, \int_{\cI^+_-} d\O_{D-2} \; \e^{\p \big(-\tfrac{p}{2}-k,0\big)} \D H_{ur}^{\p \big(\tfrac{4-p}{2}+k,0\big)}\,.
\end{equation}

\section{Algebra of charges}\label{sec5}
In our problem, the only sector of the gauge parameter that enters the charges is the scalar 
$\epsilon'$ in the Hodge decomposition on the celestial sphere. The renormalized charges 
organize into a tower of $N+1$ contributions 
$\{Q^{(0)},\dots,Q^{(N)}\}$. Their universal form is
\begin{equation}
Q_{B} \;=\; \sum_{k=0}^{N} Q^{(k)} \,, \qquad
Q^{(k)} \;=\; \int_{\mathcal I^+_-} d\Omega_{D-2}\, 
\epsilon^{\p\left(-\tfrac{p}{2}-k,\,0\right)}(\{x^i\}) \;
\Delta \, H_{ur}^{\p\left(\tfrac{4-p}{2}+k,\,0\right)}(\{x^i\}) \, , 
\label{eq:charge-tower}
\end{equation}
The expression follows from the covariant definition of the charge together with the 
renormalization procedure that removes divergences.

We define the bracket between two charges via the presymplectic form $\Omega$ on the 
solution space, imposing $i_{\epsilon}\Omega = \delta Q[\epsilon]$. Then, for two 
parameters $\epsilon,\eta$, the bracket reads
\begin{equation}
\{ Q[\epsilon] , Q[\eta] \} \;:=\; \delta_{\eta} Q[\epsilon] 
\;=\; \int_{\mathcal I^+_-} d\Omega_{D-2}\, 
\epsilon'\, \Delta \, (\delta_{\eta} H'_{ur}) \, , 
\label{eq:bracket-def}
\end{equation}
since $\epsilon'$ is an independent parameter that does not vary under $\delta_\eta$  
and the measure is invariant. The field strength, $H=dB$ is gauge-invariant and so every components; furthermore the component $H_{uri_1...i_{p-1}}$ admits the Hodge decomposition \eqref{hodge} which in coordinates reads as in \eqref{hodgeex}. Imposing gauge invariance returns that $H'_{ur}$ is also gauge invariant while the gauge transformation of the form appearing in the exact part coincide with the exterior derivative of the gauge parameter.
Therefore, using \eqref{eq:charge-tower} we get
\begin{equation}
0 \;=\; \{ Q_B[\epsilon], Q_B[\eta] \}
 \;=\; \left\{ \sum_{k=0}^{N} Q_B^{(k)}[\epsilon], \, 
              \sum_{k'=0}^{N} Q_B^{(k')}[\eta] \right\}
 \;=\; \sum_{k,k'=0}^{N} \{ Q_B^{(k)}[\epsilon], \, Q_B^{(k')}[\eta] \} \, .
\end{equation}
Therefore, the global identity translates into a hierarchy of conditions 
among the different orders \( k, k' \).
Since we are expanding the charges in a power series we can collect terms by degree
\begin{equation}
0 \;=\; \sum_{k''=0}^{2N} \; \sum_{\substack{k+k'=k''}} 
\{ Q_B^{(k)}[\epsilon], \, Q_B^{(k')}[\eta] \} \, .
\end{equation}
This means that, for each degree \( k'' \), we have the condition
\begin{equation}
\sum_{\substack{k+k'=k''}} \{ Q_B^{(k)}[\epsilon], \, Q_B^{(k')}[\eta] \} \;=\; 0 \, .
\end{equation}
Therefore, asymptotic symmetries do not indefinitely multiply the number of degrees of freedom; rather, they organize themselves into a tower of modes that are interconnected. Moreover, the abelian property is the analogue, for the $O(r^N)$ tower, of the well-known abelian 
structure of large gauge symmetries in electrodynamics, now extended to $p$-forms and to 
the full hierarchy in $r$. Possible central extensions of the charge algebra can arise due to two mechanisms:
\begin{itemize}
    \item magnetic/dual sectors. There always exist two types of charges: the electric ones and the magnetic ones. Considering also the magnetic sector or mixed electric/magnetic sectors, can produce central extensions in the charge algebra. Thus, even if the underlying group is abelian, the set of charges \( \{ Q_E[\epsilon], Q_M[\eta] \} \) may realize a central extension, as in the classical Kac--Moody cases. The magnetic charges can be constructed following the same reasoning of the electric ones, schematically
    \begin{equation}
        \tilde{Q}_B \propto \int_{\mathcal{I}^+_-} d\Omega_p \,
\tilde{\epsilon} \,
\Delta\!\Big[
  (\star d\phi)_{r i_1 \cdots i_p} \,
  \gamma^{i_1 j_1} \cdots \gamma^{i_p j_p} \,
  \epsilon_{j_1 \cdots j_p}
\Big] \, .
    \end{equation}
    where the magnetic "gauge" parameter is nothing but a harmonic function on the celestial sphere in order to not spoil the chosen fall-offs. In the mixed electric/magnetic sector, the gauge variation is not necessarily zero, as it was instead for the purely electric sector, which implies the presence of a central extension in the case in which the variation does not depend on the fields and a field-dependent extension otherwise;
    \item Renormalization ambiguities. The presymplectic formalism admits local ambiguities that are fixed fixed by imposing renormalization conditions. 
However, they are not uniquely determined and different choices lead to charge definitions 
that can differ by surface terms. If two charges \( Q[\epsilon] \) and \( Q[\eta] \) differ by non-trivial surface terms, their bracket may acquire a constant contribution realizing a central extension. Since we have not used the finite term \( \Xi^{(0)} \) in the renormalization procedure, this could be lead to possible central extension of the charge algebra.
\end{itemize}

%%%%%%%%%%%%%%%%%%%%%%%%%%%%%%%%%%%%%%%%%%%%%%%%%%%%%%%%%%%%%%%%%%%%%%%%%%%%%%%%%%%%%%%%%%%%%%%%%%%%%%%%%%%%%%%%%%%%%%%%%%%%%%%%%%%%%%%%%%%%%%%%%%%%%%%%%%%%%%%%%%%%%%%%%%%%%%%%%%%%%%%%%%%%%%%%%%%%%%%%%

\section{Conclusions}
In this work, we studied higher-order asymptotic symmetries for $p$-forms in $D=p+2$, which are on-shell duals to a scalar. Indeed, by means of the Hodge decomposition, we were able to reduce the complexity of the problem to a scalar one, expressing the asymptotic charges in terms of a scalar parameter times a $ur$ component of a field strength. Surprisingly, up to some overall factors this problem turns out to be $p$-independent. Furthermore, let us remark that, from a dual perspective, this is the first derivation of any scalar asymptotic charges in an odd spacetime dimension and of higher-order scalar asymptotic charges in any spacetime dimension different from $D=4$. In fact, by means of symplectic renormalization, we were able to understand the divergences of the asymptotic charges in terms of ambiguities of the presymplectic potential. Consistently with the other analyses in Lorenz gauge, both $t$- and $u$-divergences were addressed. Along the way, we have shown that the presymplectic potential admits a general structure independent of the specific $p$-form theory. This could also be valid in the case of mixed symmetry tensors and it is a possible direction for the future. 

This approach suggests the possibility of discovering an infinite set of asymptotic charges, broadly meaning an infinite number of conserved charges. We stress, however, two points. First, as noted in \cite{manz2}, in sufficiently high dimension (surely greater than $D=p+2$) it seems to be possible to perform a gauge-for-gauge fixing to cut off the series of the pure gauge terms to start with the order $\cO\left(r^{{(D-2p-2)}/{2}}\right)$. However, the dimension needed to perform this gauge-for-gauge fixing grows proportionally with the (higher) order of the gauge parameter expansion. To derive this result, it is necessary the use of a theorem that characterises the solutions of the harmonic-divgrad system (a system of PDEs needed to discuss gauge-for-gauge fixing) on the sphere \cite{manz5}. Second, it is crucial to emphasise how we used a specific prescription: to cancel the divergences by means of ambiguities while leaving unaltered the finite parts, which enter in the definition of asymptotic charges. However, it is clear that there are residual ambiguities that might further constraint the number of physical asymptotic charges. In particular, we note that we did not use the residual term $\delta \Xi_{(0)}^t$. In this perspective could be possible, appropriately modifying the procedure, to renormalize the divergent asymptotic charges appearing in \cite{manz2} associated to Coulomb falloffs of the field. Furthermore, we compute the charge algebra which results in an abelian structure generalizing the case of Maxwell fields. The possibility of central extensions lies in the mixed electric/magnetic sector and/or due to the ambiguities of the presymplectic potential. 

An enticing perspective regards the study of higher-order asymptotic symmetries in gravity, for which a full $\cO(r^N)$ analysis is still missing. We seek to analyse also theories with higher-spin and mixed-symmetry tensor. Moreover, as $p$-forms are part of the string spectrum and $p$-form gauge theories emerge in the tensionless limit of string theory, the calculation of asymptotic charges and their renomalization may shed new light on the infrared limit of string theory and $D_p$-branes.

In addition to this, we would like to explore better the connection with soft theorems, which are not fully understood for $p$-forms. In this sense, the analysis for the duality with a scalar is appealing, since scalar theories represent the simplest models in physics.  This connection could also help in understanding the potential role of the residual ambiguities that are left in the symplectic renormalization procedure, whose role is currently under several investigations. 

\appendix
	\section{Equations of motion}
    \subsection{Gauge fields}
	Let us derive the equation of motion for a $p$-form gauge field in a $D=p+2$-dimensional spacetime in Lorenz gauge. We have
	\begin{subequations}
		\begin{align}
			\Box \cB_{uri_1\dots i_{p-2}} 
			=& \left[ \pr_r^2 - 2\pr_u\pr_r + \frac{p-4}{r}(\pr_u  -\pr_r)+\frac{\D-p}{r^2}\right]\cB_{uri_1\dots i_{p-2}}  \notag\\
			&-\frac{2}{r^3}D^i \cB_{ui\, i_1\dots i_{p-2}}\,,
			\\
			\Box \cB_{ui_1\dots i_{p-1}} 
			=& \left[ \pr_r^2 - 2\pr_u\pr_r+\frac{p-2}{r}(\pr_u-\pr_r) +\frac{\D-(p-1)^2}{r^2} \right]\cB_{ui_1\dots i_{p-1}} \notag\\
			&\;+ \frac{2}{r}\sum_{n=1}^{p-1} D_{i_n} \cB_{u i_1 \dots i_{n-1}\,r\, i_{n+1}\dots i_{p-1}}\,,
			\\
			\Box \cB_{ri_1\dots i_{p-1}} 
			=& \left[ \pr_r^2 - 2\pr_u\pr_r+\frac{p-2}{r}(\pr_u-\pr_r) + \frac{\D-1}{r^2} \right]\cB_{ri_1\dots i_{p-1}} \notag\\
			&\;-\frac{p-2}{r^2}{\cB_{ui_1\dots i_{p-1}}}-\frac{2}{r^3} D^i \cB_{ii_1\dots i_{p-1}}- \frac{2}{r}\sum_{n=1}^{p-1} D_{i_n} \cB_{r i_1 \dots i_{n-1}\,u\, i_{n+1}\dots i_{p-1}}\,,
			\\
			\Box \cB_{i_1\dots i_{p}} 
			=& \left[ \pr_r^2 - 2\pr_u\pr_r+\frac{p}{r}(\pr_u-\pr_r) + \frac{\D}{r^2} \right]\cB_{i_1\dots i_{p}} \notag\\
			&\;- \frac{2}{r}\sum_{n=1}^{p} D_{i_n} (\cB_{i_1 \dots i_{n-1}\,u\, i_{n+1}\dots i_{p}}-\cB_{i_1 \dots i_{n-1}\,r\, i_{n+1}\dots i_{p}})\,,
		\end{align}
	\end{subequations}
	while the gauge condition $ G_{{\m_1}\dots{\m_{p-1}}}:=\nabla^\m\cB_{\m \m_1 \dots \m_{p-1}}$ reads
	\begin{subequations}
		\begin{align}
			&G_{ur i_1\dots i_{p-3}}=\frac{1}{r^2}D^i B_{urii_1\dots i_{p-3}}\,,
			\\
			&G_{u i_1\dots i_{p-2}} = \left(\pr_u-\pr_r +\frac{p-4}{r}\right) B_{uri_1\dots i_{p-2}}-\frac{1}{r^2}D^i B_{u\,i\,i_1\dots i_{p-2}}\,,
			\\
			&G_{r i_1\dots i_{p-2}}=\left(-\pr_r +\frac{p-4}{r}\right)\cB_{u r i_1\dots i_{p-2}}-\frac{1}{r^2}D^iB_{ri\,i_1\dots i_{p-2}}\,,
			\\
			&G_{i_1\dots i_{p-1}}=\frac{1}{r^2}D^i \cB_{i\,i_1\dots i_{p-1}}-\pr_u \cB_{ri_1\dots i_{p-1}} - \left( \pr_r - \frac{p-2}{r} \right)(\cB_{ui_1\dots i_{p-1}}-\cB_{ri_1\dots i_{p-1}})\,.
		\end{align}
	\end{subequations}
	Order by order, the equations for the physical components $B$ of the gauge fields, which admit a power expansion, are
	\begin{subequations}
		\begin{align}
            \notag(-2n-p+6)\pr_uB_{uri_1\dots i_{p-2}}^{(n-1)} 
			=&\big[ \D +(n-2)(n+p-5)  -p \big]B_{uri_1\dots i_{p-2}}^{(n-2)}\\
			&\; -2 D^i B_{uii_1\dots i_{p-2}}^{(n-3)}\,,
			 \\ \notag
			(-2n-p+4)\pr_uB_{ui_1\dots i_{p-1}}^{(n-1)} =&\big[ \D +(n-2)(n+p-3)  -(p-1)^2 \big]B_{ui_1\dots i_{p-1}}^{(n-2)}\\
			&\; +2 \sum_{k=1}^{p-1}D_{i_k} B_{ui_1\dots r\dots i_{p-1}}^{(n-1)}\,,
			\\\notag
			(-2n-p+4)\pr_uB_{ri_1\dots i_{p-1}}^{(n-1)} =&\big[ \D +(n-2)(n+p-3)  -1\big]B_{ri_1\dots i_{p-1}}^{(n-2)} -(p-2)B_{ui_1 \dots i_{p-1}}^{(n-2)} \\
			&\; -2D^iB_{i i_1\dots i_{p-1}}^{(n-3)}+2 \sum_{k=1}^{p-1}D_{i_k} B_{ui_1\dots r\dots i_{p-1}}^{(n-1)}\,,
			\\ \notag
			(-2n-p+2)\pr_uB_{i_1\dots i_{p}}^{(n-1)} =&\big[ \D +(n-2)(n+p-1)  \big]B_{i_1\dots i_{p}}^{(n-2)}\\
			&-2\sum_{k=1}^p D_{i_k} (B_{i_1 \dots i_{k-1}\,u\, i_{k+1}\dots i_{p}}^{(n-1)}-B_{i_1 \dots i_{k-1}\,r\, i_{k+1}\dots i_{p}}^{(n-1)}) \,.
		\end{align}
	\end{subequations}
	The order-by-order gauge condition reads 
	\begin{subequations}
		\begin{align}
			&D^i B_{urii_1\dots i_{p-3}}^{(n)}=0 
			\\
			& \pr_uB_{uri_1\dots i_{p-2}}^{(n)} =- (n+p-5)B_{uri_1\dots i_{p-2}}^{(n-1)}+D^i B_{ui_1\dots i_{p-1}}^{(n-2)}
			\\
			&(n+p-4)B_{uri_1\dots i_{p-2}}^{(n)}=D^i B_{ri_1\dots i_{p-1}}^{(n-1)}
			\\
			&\pr_uB_{ri_1\dots i_{p-1}}^{(n)} =D^i B_{i i_1 \dots i_{p-1}}^{(n-2)}+(n+p-3)\big( B_{ui_1\dots i_{p-1}}^{(n-1)}- B_{ri_1\dots i_{p-1}}^{(n-1)}\big)
		\end{align}
	\end{subequations}

	In the asymptotic charge, only $\e_{i_1\dots i_p}$ is relevant and, specifically, the only contribution comes from the $\e^\p$ component in \eqref{Hodge_par}. Interestingly, that component completely decouples from the rest of the equations, reducing our problem to a scalar one, with the corresponding system being 
    	\begin{subequations}\label{gauge_par_eom}
		\begin{align}
			&(-2n-p+4)\pr_u\he^{\p \,(n-1)} =\big[ \D +(n-2)(n+p-3) \big]\he^{\p\,(n-2)}\\
			\notag&(-2n-p+4)\pr_u\e^{\p \,(n-1)} +2\pr_u\he^{\p \,(n-1)}=\big[ \D +(n-2)(n+p-3) \big]\e^{\p\,(n-2)}\\
			&\qquad \qquad \qquad \qquad \qquad \qquad \qquad \qquad \quad -(2n+p-3) \he^{\p\,(n-2)}
		\end{align}
	\end{subequations}
	where we used the commutation relation
	\begin{equation}
		\D D^i \vf = D^i (\D+D-3)\vf\,. 
	\end{equation}
	The transversality condition does not contain $\e^\p$ and therefore we do not need to discuss it. In determining the relevant field-strength component, we also need this commutation relation
    \begin{equation}
        \begin{split}
            \D D^i B_{ri i_1 \dots i_{p-2}}         &=D^i (\D -D+2p-1 )B_{ri i_1 \dots i_{p-2}} \,.
        \end{split}
    \end{equation}

	\section{Solutions of the EOM}\label{AppB}
	\subsection{Falloffs}
	We start with the standard radiation falloffs in $D=p+2$
	\begin{equation}
			B_{i_1\dots i_p} = \cO\left(r^{\tfrac{p}{2}}\right), \;\;  B_{ri_1\dots i_{p-1}}  \,, \,
			B_{ui_1\dots i_{p-1}}  = \cO\left(r^{\tfrac{p-2}{2}}\right),\quad  B_{uri_1\dots i_{p-2}}  = \cO\left(r^{\tfrac{p-4}{2}}\right)\,.
	\end{equation}
	To leading order, the wave equation does not bring about any constraint, leaving all the leading terms with arbitrary angular dependence. The gauge condition, implies that
	\begin{equation}
		\begin{split}
			&D^i B_{urii_1\dots i_{p-3}}^{\big(\tfrac{4-p}{2}\big)}=0 \,,
			\\
			& \pr_uB_{uri_1\dots i_{p-2}}^{\big(\tfrac{2-p}{2}\big)} =0\,,
			\\
			& (p-4)B_{uri_1\dots i_{p-2}}^{\big(\tfrac{4-p}{2}\big)}=2D^i B_{ri_1\dots i_{p-1}}^{\big(\tfrac{2-p}{2}\big)}\,,
			\\
			&\pr_uB_{ri_1\dots i_{p-1}}^{\big(\tfrac{2-p}{2}\big)} =0\,.
		\end{split}
	\end{equation}	
	When studying the next-to-leading order equations, we find two constraints. In particular, the equations of motion give 
	\begin{equation}
		\begin{split}
			-2\pr_uB_{uri_1\dots i_{p-2}}^{\big(\tfrac{6-p}{2}\big)} 
			=&\left[ \D -\frac{p^2-2p+8}{4}\right]B_{uri_1\dots i_{p-2}}^{\big(\tfrac{4-p}{2}\big)} -2 D^i B_{uii_1\dots i_{p-2}}^{\big(\tfrac{2-p}{2}\big)}\,,
			\\
			-2\pr_uB_{ri_1\dots i_{p-1}}^{\big(\tfrac{4-p}{2}\big)} =&\left[ \D -\frac{p^2-2p-4}{4}\right]B_{ri_1\dots i_{p-1}}^{\big(\tfrac{2-p}{2}\big)} -2D^iB_{i i_1\dots i_{p-1}}^{\big(-\tfrac{p}{2}\big)}\\
			&-(p-2)B_{ui_1 \dots i_{p-1}}^{\big(\tfrac{2-p}{2}\big)}+2 \sum_{k=1}^{p-1}D_{i_k} B_{ui_1\dots r\dots i_{p-1}}^{\big(\tfrac{4-p}{2}\big)}\,.
		\end{split}
	\end{equation}
	From the gauge condition, we read
	\begin{equation}
		\begin{split}
			& \pr_uB_{uri_1\dots i_{p-2}}^{\big(\tfrac{6-p}{2}\big)} =\frac{4-p}{2}B_{uri_1\dots i_{p-2}}^{\big(\tfrac{4-p}{2}\big)}+D^i B_{ui_1\dots i_{p-1}}^{\big(\tfrac{2-p}{2}\big)}\,,
			\\
			&(n+p-4)B_{uri_1\dots i_{p-2}}^{(n)}=D^i B_{ri_1\dots i_{p-1}}^{(n-1)}\,,
			\\
			&\pr_uB_{ri_1\dots i_{p-1}}^{\big(\tfrac{4-p}{2}\big)} =D^i B_{i i_1 \dots i_{p-1}}^{\big(-\tfrac{p}{2}\big)}+\frac{p-2}{2}\big( B_{ui_1\dots i_{p-1}}^{\big(\tfrac{2-p}{2}\big)}- B_{ri_1\dots i_{p-1}}^{\big(\tfrac{2-p}{2}\big)}\big)\,.
		\end{split}
	\end{equation}
	From the first equations of each set, we find 
	\begin{equation}
		\left[ \D -\frac{p(p-2)}{4}\right]B_{uri_1\dots i_{p-2}}^{\big(\tfrac{D-2p+2}{2}\big)} =0\,,
	\end{equation}
	which implies that the leading order $B_{uri_1\dots i_{p-2}}^{\big(\tfrac{4-p}{2}\big)}$ does not have an arbitrary angular dependence and, specifically, it is a constant term for $D=4$ and there is no solution for $D>4$, since the two-sphere Laplacian is negative semidefinite. Hence, we consistently consider 
	\begin{equation}
		B_{uri_1\dots i_{p-2}}^{\big(\tfrac{4-p}{2}\big)}=0\,.
	\end{equation}
	Substituting this condition, we find
	\begin{equation}
		D^i B_{rii_1\dots i_{p-2}}^{\big(\tfrac{2-p}{2}\big)}=0\,.
	\end{equation}
	
	Performing the same calculation for the second equations of each set, we find
	\begin{equation}
		\left[ \D -\frac{p^2+2p-4}{4}\right]B_{ri_1\dots i_{p-1}}^{\big(\tfrac{2-p}{2}\big)}=0\,,
	\end{equation}
	which tells us that 
	\begin{equation}\label{falloff_Bri}
		\begin{split}
			&B_{rii_1\dots i_{p-2}}^{\big(\tfrac{2-p}{2}\big)}=0\,.
		\end{split}
	\end{equation}
The correct falloff in Lorenz gauge are thus
\begin{equation}
		\begin{split} 
			&B_{i_1\dots i_p} = \cO\left(r^{\tfrac{p}{2}}\right)\,, \;\,\quad\qquad  B_{ri_1\dots i_{p-1}}  = \cO\left(r^{\tfrac{p-4}{2}}\right)\,, \\
			&B_{ui_1\dots i_{p-1}}  = \cO\left(r^{\tfrac{p-2}{2}}\right)\,,\qquad  B_{uri_1\dots i_{p-2}}  = \cO\left(r^{\tfrac{p-6}{2}}\right)\,.
		\end{split}
	\end{equation}

	\subsection{Boundary small gauge fixing and field strength}
    Our aim is to determine the order-by-order expression of the relevant field strength component in terms of the Cauchy data. To achieve it, we perform a further gauge fixing, employing some small residual gauge freedom. The relevant component of the field strength is 
    \begin{equation}
		H_{ur i_1\dots i_{p-1}}^{(n)} = \pr_u  B_{r i_1\dots i_{p-1}}^{(n)} +(-1)^{p}(n-1)B_{u i_1\dots i_{p-1}}^{(n-1)} + \sum_{n=1}^{p-1} D_{i_n} B_{i_{n+1}\dots uri_1 \dots i_{n-1}}^{(n)}\,.
	\end{equation}
    Specifically, the quantity that enters in the charge is
     \begin{equation}
        H_{ur}^{\p(n)} = \pr_u B_{r }^{\p(n)} +(-1)^{p}(n-1)B_{u }^{\p(n-1)}\,,
    \end{equation}
    and, from the gauge condition, we find
    \begin{equation}\label{gaugeB'r}
        \pr_u B_{r }^{\p(n)} = (-1)^{p-1}B^{(n-2)} + (n+p-3) (B_{u }^{\p(n-1)} -B_{r }^{\p(n-1)}   )\,,
    \end{equation}
    which gives
     \begin{equation}
        H_{ur}^{\p(n)} = (-1)^{p-1}B^{(n-2)} + (n+p-3) (B_{u }^{\p(n-1)} -B_{r }^{\p(n-1)}   ) +(-1)^{p}(n-1)B_{u }^{\p(n-1)}\,,
    \end{equation}
    At the same time, from the equations of motion we read
    \begin{equation}
        (-2n-p+2)\pr_uB_{r}^{\p(n)} =\big[ \D +n(n+p-3)\big]B_{r}^{\p(n-1)} -(p-2)B_u^{\p(n-1)} + (-1)^{p}2B^{(n-2)}\,.
    \end{equation}
    Combining this with \eqref{gaugeB'r}, we obtain
    \begin{equation}
       \begin{split} 
       &\left[ \D-(n+p-1)(n+p-2) \right]B_{r}^{\p(n)} +(-1)^{p-1}{(2n+p-2)}B^{(n-1)}\\
       &+(n + p - 1) (2 n + p -2)B_{u }^{\p(n)}=0\,.
       \end{split}
    \end{equation}
   For $n= \tfrac{2-p}{2}$, this equations gives exactly \eqref{falloff_Bri}. For $n> \tfrac{2-p}{2} $, we obtain
\begin{equation}
\begin{split}
     B_{u}^{\p(n)}=&\frac{(-1)^{p}}{n+p-1}B^{(n-1)}-\left[ \frac{\D}{(n + p - 1) (2 n + p -2)}-\frac{n+p-2}{2n+p-1} \right]B_{r}^{\p(n)}\,.
     \end{split}
\end{equation}
The asymptotic charge is given by 
\begin{equation}\label{cahrge}
    Q_B = \sum_{k=0}^N \, \int_{\cI^+_-} d\O_{D-2} \; \e^{\p \big(-\tfrac{p}{2}-k,0\big)} \D H_{ur}^{\p \big(\tfrac{4-p}{2}+k,0\big)}\,.
\end{equation}
In order to express $H_{ur}^{\p(n,0)}$ in terms of the standard Cauchy data $B$, we perform a small gauge fixing at the boundary
\begin{equation}\label{Boundary_gauge_fixing}
    B^{\p(n,0)}_r=0\,,
\end{equation}
for all $n\geq \tfrac{4-p}{2}$ using the free terms of the subleading gauge parameters $\e^{(n,0)}$ with $n>\tfrac{4-p}{2}$ and $n\neq 0$. The gauge transformation of $B^{\p(n,0)}_r$ depends only on the $\ep$ part of the Hodge decomposition of the gauge parameter. Indeed by using the Hodge decompositions of $B_{ri_1...i_{p-1}}$ and $\e_{i_1...i_{p-1}}$, by comparing with the terms in $\delta_{\e}B_{ri_1...i_{p-1}}$ we get 
\begin{equation}\label{gaugevar}
    \delta_{\e}B^{\p(n,0)}_r=-(n-1)\e^{\p(n-1,0)} \, . 
\end{equation}
The parameters appearing in \eqref{gaugevar} are subleading with those in the charge since $\frac{2-p}{2}+k > -\frac{p}{2}-k$ for every $k\geq 0$. Therefore they do not contribute to the charge \eqref{cahrge} and are associated to a vanishing charge. The gauge fixing \eqref{Boundary_gauge_fixing} can be achieved for every $n$ except $n=1$, as observed also in \cite{Romoli:2024hlc}, which occurs only for even $p$. In practice, we cannot gauge away $B^{\p(1,0)}_r$, which is the natural Cauchy data at that specific order. However, we can gauge away $B^{(0,0)}$ using the residual free $\e^{(0,0)}$ gauge parameter. 

In the end, we find
 \begin{equation}
       \begin{split}
            H_{ur}^{\p\big(\tfrac{4-p}{2},0\big)} &= (-1)^{p-1} B^{(-{p}/{2},0)} \,,\\
             H_{ur}^{\p(2,0)} \quad\;&= \frac{\D}{p} B_r^{\p(1,0)}\,\,\,\, \qquad\qquad\qquad\qquad \ \text{if} \  p \in 2\mathbb{N}\,,\\
           H_{ur}^{\p(n,0)} \quad\; &=  \frac{(-1)^{p-1}+n-1}{n+p-2} B^{(n-2,0)} \qquad n\neq \tfrac{4-p}{2} \; ,
       \end{split} 
    \end{equation}
    where the second equation holds only for even $p$.

\section*{Conflict of Interest, Funding and Data Availability}
The authors declare no conflict of interest. This research received no external funding. No new data were created or analyzed in this study.

	\bibliographystyle{JHEP} 
	
	\bibliography{main}
	
\end{document}